# PUBLIC CONSTITUTIONAL AI

*Gilad Abiri*[*]

*We are increasingly subjected to the power of AI authorities. Machine learning models now underpin algorithmic markets, determine whose speech is amplified or restricted, shape government decisions ranging from resource allocation to predictive policing, and influence our access to information on critical issues such as voting and public health. As AI decisions become inescapable, entering domains such as healthcare, education, and law, we must confront a vital question: How can we ensure that AI systems, which increasingly regulate our lives and make decisions that shape our societies, have the authority and legitimacy necessary for effective governance?*

*To secure AI legitimacy, we need to develop methods that engage the public in the project of designing and constraining AI systems, thereby ensuring that these technologies reflect the shared values and political will of the communities they serve. Constitutional AI, proposed and developed by Anthropic AI, represents a step towards this goal, offering a model for how AI might be brought under democratic control and made answerable to the common good.*

*Just as constitutions limit and guide the exercise of governmental power, Constitutional AI seeks to hardcode explicit principles and values into AI models, rendering their decisionmaking more transparent and accountable. What sets Constitutional AI apart is its commitment to grounding AI training in a clear, human-understandable "constitution." By training AI to adhere to principles legible to both humans and machines, this approach aims to foster trust and stability in the development of these increasingly powerful technologies.*

*However, I argue that Constitutional AI, in its current form (developed by a private corporation seeking to create*

[*] Associate Professor of Law, Peking University School of Transnational Law; Affiliate Faculty, Information Society Project, Yale Law School; and Senior Research Affiliate, Singapore Management University Centre for Digital Law. The author wishes to thank Li Kejun, Wang Xi, and Yin Xinyi for exceptional research assistance.

*universally applicable constitutional principles), is unlikely to fully resolve the crisis of AI legitimacy due to two key deficits: First, the opacity deficit, which suggests that the inherent complexity of AI systems undermines our ability to reason out their decisionmaking. Second, the political community deficit, which suggests that AI systems are grounded in abstract models rather than in human judgment, lacks the social context that legitimizes authority.*

*To remedy these deficits, I propose Public Constitutional AI, a framework that involves the public in drafting an AI constitution that must be used in the training of all frontier AI models operating within a given jurisdiction. By transforming the AI constitution from a technical solution devised by engineers into a product of significant citizen involvement, Public Constitutional AI mitigates the opacity deficit. It does so by rendering the principles and values governing AI systems more transparent and accessible to the forms of public discourse and contestation essential to democratic legitimacy. Moreover, by grounding the development of AI principles in the social context and shared experiences of a particular political community, Public Constitutional AI helps bridge the gap between the abstract logic of algorithms and the situated, contextual judgments that legitimize authority in a democracy, thereby mitigating the political community deficit.*

## I. INTRODUCTION

The rapid rise of algorithmic decisionmaking is transforming both private and public spheres. From content moderation to criminal sentencing, AI systems are increasingly entrusted with choices that carry profound consequences for individuals and society.[1] As machine learning and big data become the backbone of vital government functions and shape the information ecosystem, concerns about the legitimacy of these AI authorities[2] are growing.[3]

The increasingly expanding reach of AI authorities raises fundamental questions about the legitimacy of power in the digital age. Legal and political theorists have long recognized that stable and effective governance requires more than mere coercion; it demands legitimacy—the widespread belief that power is being exercised in a rightful manner.[4] Without this perception of justified

---

[1] *See* Danielle Keats Citron, *Technological Due Process*, 85 WASH. U. L. REV. 1249, 1253–54, 1257 (2008) (highlighting concerns regarding the influence of AI systems on criminal investigations and "the adjudication of important individual rights"); *see also* TARLETON GILLESPIE, CUSTODIANS OF THE INTERNET: PLATFORMS, CONTENT MODERATION, AND THE HIDDEN DECISIONS THAT SHAPE SOCIAL MEDIA 22 (2018) ("[Social media platforms] are designed so as to invite and shape our participation toward particular ends. This includes how profiles and interactions are structured; how social exchanges are preserved; how access is priced or paid for; and how information is organized algorithmically, privileging some content over others, in opaque ways.").

[2] I will use the expressions "automated authorities" and "AI authorities" interchangeably.

[3] *See* Ryan Calo & Danielle Keats Citron, *The Automated Administrative State: A Crisis of Legitimacy*, 70 EMORY L.J. 797, 800–03 (2021) (discussing transparency, accountability, and due process concerns arising from the use of AI systems by federal and state agencies to automate decisionmaking); *see also* DAVID FREEMAN ENGSTROM, DANIEL E. HO, CATHERINE M. SHARKEY & MARIANO-FLORENTINO CUÉLLAR, ADMIN. CONF. U.S., GOVERNMENT BY ALGORITHM: ARTIFICIAL INTELLIGENCE IN FEDERAL ADMINISTRATIVE AGENCIES 7 (2020) ("When public officials deny benefits or make decisions affecting the public's rights, the law generally requires them to explain why. Yet many of the more advanced AI tools are not, by their structure, fully explainable. A crucial question will be how to subject such tools to *meaningful accountability* and thus ensure their fidelity to legal norms of transparency, reason-giving, and non-discrimination.").

[4] *See* Richard H. Fallon, Jr., *Legitimacy and the Constitution*, 118 HARV. L. REV. 1787, 1795 (2005) ("[L]egitimacy signifies an active belief by citizens, whether warranted or not, that particular claims to authority deserve respect or obedience for reasons not restricted to self-interest."); Tom R. Tyler, *Psychological Perspectives on Legitimacy and Legitimation*, 57 ANN. REV. PSYCH. 375, 376–77 (2006) ("[U]nder some circumstances people are also influenced by others because they believe that the decisions made and rules enacted by others are in some way right or proper and ought to be followed. In other words, subordinates also 'relate to the

authority, compliance becomes fragile, dependent on constant surveillance and the threat of force.[5] In the context of algorithmic decisionmaking, the opacity and inscrutability of advanced AI systems pose significant challenges to establishing their legitimacy in the eyes of those subject to their power.[6]

My goal in this essay is to engage with one potential solution to the issue of AI legitimacy: Constitutional AI. This idea and technology, developed by Anthropic AI—the team behind Claude, one of the frontier large language models (LLMs)[7]—attempts to hardcode a document containing explicit principles and values into AI systems, analogous to how constitutions operate to structure and constrain government authority.[8]

By training AI models to adhere to principles accessible to both humans and models, Constitutional AI aims to make their decisionmaking more transparent, accountable, and aligned with human values.[9] However, I argue that while private efforts like

---

powerful as moral agents as well as self-interested actors; they are cooperative and obedient on grounds of legitimacy as well as reasons of prudence and advantage.'" (citations omitted)); *see also* Gilad Abiri & Sebastián Guidi, *From a Network to a Dilemma: The Legitimacy of Social Media*, 26 STAN. TECH. L. REV. 92, 107 (2023) ("Legitimation occurs when a powerholder behaves according to the beliefs people have about the rightful way of exercising authority.").

[5] *See* DAVID BEETHAM, THE LEGITIMATION OF POWER 27–28 (Peter Jones & Albert Weale eds., 1991). ("[W]hen legitimacy is eroded or absent. . . . coercion has to be much more extensive and omnipresent, and that is costly to maintain. Moreover, the system of power has only one line of defense, that of force; and it can therefore collapse very rapidly . . . .").

[6] *Cf.* Simon Chesterman, *Through a Glass Darkly: Artificial Intelligence and the Problem of Opacity*, 69 AM. J. COMPAR. L. 271, 273–74 (2021) ("Distinct from the challenges posed by autonomy in A.I. systems, the increasing opacity of those systems is not a challenge to the centrality of human agents as legal *actors* so much as to our ability to understand and evaluate *actions*—something essential to meaningful regulation." (footnote omitted)).

[7] ANTHROPIC, https://www.anthropic.com/company (last visited Apr. 2, 2025).

[8] *Cf.* Jenna Burrell, *How the Machine 'Thinks': Understanding Opacity in Machine Learning Algorithms*, BIG DATA & SOC'Y, Jan.–June 2016, at 1, 9 ("Finding ways to reveal something of the internal logic of an algorithm can address concerns about lack of 'fairness' and discriminatory effects, sometimes with reassuring evidence of the algorithm's objectivity . . . ."); Andrew D. Selbst & Solon Barocas, *The Intuitive Appeal of Explainable Machines*, 87 FORDHAM L. REV. 1085, 1089 (2018) ("Where intuition is insufficient to determine whether the model's rules are reasonable or rest on valid relationships, justification can sometimes be achieved by demonstrating and documenting due care and thoughtfulness.").

[9] *See* Yuntao Bai et al., *Constitutional AI: Harmlessness from AI Feedback*, ARXIV 1–2, 5 (Dec. 15, 2022, 6:19 AM), https://arxiv.org/pdf/2212.08073 [https://perma.cc/KUF4-FHD5] (describing experimentation with methods for training a "harmless AI assistant" through a

Anthropic's are a promising start, they will ultimately fall short of securing robust legitimacy for AI authority. What is needed is an approach that takes the constitutional analogy seriously by engaging the public in a process of AI constitution-making. This article engages in the thought experiment of developing such a "Public Constitutional AI." I call it a thought experiment since the reader must accept two plausible but unproven premises to follow the argument: First, that AI will dramatically transform power relationships in digital societies.[10] Second, since the article makes a political and legal point, that the technological aspects of Constitutional AI, discussed below, can actually do what their developers claim.[11]

The vision of Public Constitutional AI advanced here would apply not only to AI systems operated by government entities but also to those developed and deployed by private actors. Given the increasingly influential role that private companies play in shaping the informational and communicative infrastructure of our societies,[12] subjecting their AI systems to public oversight and accountability is essential for promoting democratic legitimacy in the algorithmic age.[13] Public Constitutional AI thus represents a form of "hybrid" governance that blends public and private, recognizing the need for constitutional principles to evolve beyond the traditional state-action paradigm.[14]

---

list of rules and principles and defining this method as "Constitutional AI").

[10] This is a very widely held position. *See, e.g.*, Seth Lazar, *Automatic Authorities: Power and AI*, *in* COLLABORATIVE INTELLIGENCE: HOW HUMANS AND AI ARE TRANSFORMING OUR WORLD 37, 38 (Mira Lane & Arathi Sethumadhavan eds., 2024) ("[P]eople are increasingly subject to power exercised by means of automated systems."); *cf.* Julie E. Cohen, *Law for the Platform Economy*, 51 U.C. DAVIS L. REV. 133, 199–203 (2017) (arguing that platforms exercise quasi-governmental power and should be subject to public oversight, which applies tenfold to potential AI uses).

[11] *See* discussion *infra* Part III.A.1.

[12] *See* Kate Klonick, *The New Governors: The People, Rules, and Processes Governing Online Speech*, 131 HARV. L. REV. 1598, 1600–02 (2018) (describing how private platforms "are now essential to participation in democratic culture" and regulate online speech in ways that resemble government censorship).

[13] *See* NICOLAS P. SUZOR, LAWLESS: THE SECRET RULES THAT GOVERN OUR DIGITAL LIVES 112–114 (2019) (arguing for the application of "digital constitutionalism" to private digital platforms by "developing new guiding principles that can be applied to the unique way the internet is governed" and by increasing accountability to overcome public mistrust).

[14] *See generally* Michèle Finck, *Digital Co-Regulation: Designing a Supranational Legal*

The argument proceeds in three parts:

In Part II, I discuss the concept of AI legitimacy and identify two inherent deficits standing in the way of its realization. The first is the opacity deficit, which arises from the black box nature of advanced AI systems.[15] When the reasoning behind algorithmic decisions is inherently inscrutable, it undermines the public's ability to assess their fairness and hold power accountable.[16] This opacity operates at both the individual level, where the specific factors driving any given decision are often unknowable, and the systemic level, where the general rules and assumptions baked into the algorithm remain hidden from view.[17] The second legitimacy deficit is the political community deficit, stemming from AI's lack of grounding in any specific social context.[18] As a technology of pure statistical abstraction, AI cannot engage in the discursive processes that legitimize authority within a self-governing polity.[19] Unlike human decisionmakers, who are embedded in the shared meanings and norms of a particular community, AI operates according to the asocial logic of optimization and statistical inference.[20] This

*Framework for the Platform Economy*, 43 EUR. L. REV. 47 (2018) (proposing a co-regulatory model for the governance of digital platforms that involves both private and public actors).

[15] *See infra* Part II.B.1.

[16] *See* Cary Coglianese & David Lehr, *Transparency and Algorithmic Governance*, 71 ADMIN. L. REV. 1, 17 (2019) ("[T]he use of so-called black-box algorithms would seem, at least at first glance, to run up against the law's general requirement that government provide adequate reasons for its actions. Their use might seem to undermine basic good-government principles designed to promote accountability and build trust."); Hannah Bloch-Wehba, *Access to Algorithms*, 88 FORDHAM L. REV. 1265, 1269–70 (2020) ("As the government procures and relies upon newer, more sophisticated decision-making technologies, such as machine learning, it also makes decisions more opaque, harder to explain, and less attributable to specific causes.").

[17] *See* Selbst & Barocas, *supra* note 8, at 1089–97, 1126 (discussing the secrecy of algorithmic decisionmaking, including the critiques that the "inner workings [of algorithms] remain undisclosed," that the correlation between certain statistical relationships may be impossible to understand, and that some factors influencing decisionmaking are unobservable).

[18] *See infra* Part II.B.2.

[19] *See* Robert C. Post, *The Internet, Democracy, and Misinformation*, *in* DISINFORMATION, MISINFORMATION, AND DEMOCRACY: LEGAL APPROACHES IN COMPARATIVE CONTEXT 38, 48 (Ronald J. Krotoszynski, Jr., András Koltay & Charlotte Garden eds., 2025) ("AI cannot make content moderation decisions with the legitimacy or authority of law. . . . AI learns as it receives feedback about its decisions.").

[20] *See infra* note 129 and accompanying text.

alienation from the lived realities of citizens poses a fundamental barrier to AI's democratic legitimation.

With the challenges of AI legitimacy laid out, Part III turns to examining the idea of Constitutional AI, drawing on Anthropic's model as a key example. This approach involves codifying a set of high-level principles and values into a "constitution" that guides the behavior of their AI systems.[21] The principles attempt to reduce problematic AI responses by asking the model to reconsider prompts that are "harmful, unethical, racist, sexist, toxic, dangerous, or illegal"; "inappropriate for young children"; or "controversial or objectionable based on commonsense ethical and moral standards."[22] The AI is then trained to adhere to these principles through a process of "Constitutional AI feedback," where it is rewarded for generating outputs that align with the constitutional guidelines and penalized for deviations.[23] Over time, this process aims to instill the constitution's values into the objective function of the AI, ensuring that it will "want" to act in accordance with these principles even in novel situations.[24]

While Constitutional AI represents a well-intentioned attempt to constrain the power of artificial agents and ensure their alignment with human values, I argue that it has limited potential to address the legitimacy deficits of AI. On one hand, this approach holds promise in mitigating the systemic opacity deficit by grounding AI systems in a set of clear, accessible principles that can be scrutinized and debated by the public.[25] By making the normative foundations and constraints of AI decisionmaking more transparent and comprehensible at a system level, Constitutional AI could enhance the overall accountability of these systems and facilitate public oversight. However, Constitutional AI has little traction when it comes to addressing the opacity of specific AI decisions. Even with a transparent set of governing principles in place, the actual reasoning process behind individual determinations often remains inscrutable due to the black box nature of many AI systems.[26]

---

[21] Bai et al., *supra* note 9, at 2.

[22] *Id.* at 20–21.

[23] *See id.* at 8–12 (explaining the methods by which the AI receives feedback on its responses).

[24] *See id.* at 15 ("[W]e have moved further away from reliance on human supervision, and closer to the possibility of a self-supervised approach to alignment.").

[25] *See id.* at 4 (discussing how Constitutional AI can make the principles and constraints of AI systems more transparent and accountable).

Moreover, when it comes to the political community deficit, private Constitutional AI falls short. The unilateral and centralized nature of the constitution-drafting process, driven by a private AI corporation, fails to generate genuine political and legal legitimacy.[27] When a company like Anthropic determines the principles that will govern their AI's behavior, they are in effect telling us to trust them to make deeply consequential political choices on behalf of the broader public.[28] Without any mechanism for democratic input or deliberation, this model of Constitutional AI risks further entrenching the already significant power asymmetries between tech companies and the communities they serve.[29] The political community deficit thus remains unaddressed, as the AI system is not grounded in the shared values and deliberative processes of a self-governing polity.

---

[26] *See* discussion *infra* Part II.B.1; *see also* Cynthia Rudin, *Stop Explaining Black Box Machine Learning Models for High Stakes Decisions and Use Interpretable Models Instead*, 1 NATURE MACH. INTEL. 206, 206 (2019) ("Black box machine learning models are currently being used for high stakes decision-making throughout society, causing problems throughout healthcare, criminal justice, and in other domains. People have hoped that creating methods for explaining these black box models will alleviate some of these problems, but trying to *explain* black box models, rather than creating models that are *interpretable* in the first place, is likely to perpetuate bad practices and can potentially cause catastrophic harm to society.").

[27] *See* discussion *infra* Part III.A.2.

[28] "Ethics" language can cut both ways. On the one hand, some corporations (like Google) have employed such language "as an acceptable façade that justifies deregulation, self-regulation or market driven governance, and is increasingly identified with technology companies' self-interested adoption of appearances of ethical behavior." Elettra Bietti, *From Ethics Washing to Ethics Bashing: A View on Tech Ethics from Within Moral Philosophy*, *in* FAT* '20: PROCEEDINGS OF THE 2020 CONFERENCE ON FAIRNESS, ACCOUNTABILITY, AND TRANSPARENCY 210, 210 (2020). On the other hand, these companies may also view ethics "as a political stance which is antithetic to—instead of complementary to—serious engagement in democratic decision-making." *Id.*

[29] *See* JULIE E. COHEN, BETWEEN TRUTH AND POWER: THE LEGAL CONSTRUCTIONS OF INFORMATIONAL CAPITALISM 192 (2019) ("Mastering the processes by which technical standards are developed also requires both new kinds of regulatory expertise and new public accountability mechanisms. The language of data security, digital content management, and the like is dense and technical. It resists both public comprehension and public input, and even regulatory personnel themselves may not understand the key issues well.").

To address these shortcomings, Part IV proposes the idea of Public Constitutional AI. The idea is that to bolster the legitimacy of AI authorities, we must find ways to match the technological infrastructure of Constitutional AI with meaningful forms of public engagement and democratic deliberation. This model envisions a participatory constitution-making process where diverse stakeholders come together to deliberate on the principles that should guide AI development, regardless of whether the AI systems are deployed by public or private actors. Through public hearings, citizen assemblies, online consultations, and other mechanisms of democratic input, ordinary people would have a voice in authoring a document holding the normative foundations of AI training in their jurisdiction. The idea is that all models will be required (or given strong incentives) to be trained on the basis of the constitutional document.[30] Depending on context, the constitution-making process could be initiated and overseen by legislatures, regulatory agencies, or other public bodies with the democratic mandate to represent the interests of citizens.[31]

To complement this constitutional framework, the Public Constitutional AI approach also proposes the creation of "AI courts"—public bodies tasked with generating concrete examples and case law that illustrate how the principles of the AI constitution should be applied in practice.[32] These courts would curate a repository of paradigmatic cases and interpretations that, together with the constitutional text, would guide the training and development of new AI models. By providing both abstract principles and tangible instantiations, this approach is meant to continually engage the public

---

[30] *See* discussion *infra* Part III.A.1.

[31] *See, e.g.*, Cary Coglianese & Erik Lampmann, *Contracting for Algorithmic Accountability*, 6 ADMIN. L. REV. ACCORD 175, 197 (2021) ("We suggest that governments give greater attention to how they design and structure their contracts for services to develop and operate AI tools. Using contracting as a tool for algorithmic governance can allow governments and society to benefit from the improvements that AI tools can offer, while also helping ensure that these tools will be designed and deployed responsibly.").

[32] *See* discussion *infra* Part IV.A.2; *see also* Quan Ze Chen & Amy X. Zhang, *Case Law Grounding: Using Precedents to Align Decision-Making for Humans and AI*, ARXIV 19–22, https://arxiv.org/pdf/2310.07019 [https://perma.cc/B4GV-L3PK] (Dec. 18, 2024, 6:41 PM) (discussing a potential framework for grounding constitutional AI programming in case law).

in the process of developing the contemporary meaning of the principles of the AI Constitution.[33]

The result would encompass both the broad principles ratified by the public and the case law developed by AI courts, and it would carry the legitimating force of popular authorship by seeking to ground algorithmic power in the collective will of the community. By making AI training into a site of democratic contestation and negotiation, Public Constitutional AI could help mitigate the opacity deficit and ensure that algorithmic systems remain responsive to societal values over time. Regular opportunities for public input and oversight would enable citizens to interrogate the assumptions and value judgments embedded in AI systems, while also providing a mechanism for redress when harms or unintended consequences emerge.[34] Moreover, by fostering a sense of collective authorship of the principles governing AI, this participatory approach could help bridge the gap between the abstract logic of algorithms and the lived realities of human communities. In this way, Public Constitutional AI offers a potential path towards imbuing automated authorities with democratic legitimacy.

The social, political, and legal challenges posed by the rise of AI authority are immense, and there are no easy solutions. But by taking seriously the idea of a constitution for AI, drafted with the help of the public, we can begin to develop a path towards integrating this transformative technology into our political and legal institutions and cultures.

---

[33] *See id.* at 3 ("As a legal mechanism, case law complements constitutional (or 'statutory' law) by using concrete cases to guide decisions around concepts that are hard to concretely define otherwise. The core idea being that, by finding and adopting the right past decisions—*precedents*—we can make judgments on a new case in a way that maintains predictability against the concept while also being consistent across different decision-making agents.").

[34] *See* Natalia Díaz-Rodríguez et al., *Connecting the Dots in Trustworthy Artificial Intelligence: From AI Principles, Ethics, and Key Requirements to Responsible AI Systems and Regulation*, INFO. FUSION, June 23, 2023, at 1, 7 ("AI systems should empower human beings, allowing them to make informed decisions and fostering their fundamental rights. At the same time, proper oversight mechanisms need to be ensured, which can be achieved through human-in-the-loop, human-on-the-loop, and human-in-command approaches. In other words, AI-based systems must support human autonomy and decision making."). For an exploration of the conflicts likely to emerge at the intersection of civil rights protection and AI, see generally Sonia K. Katyal, *Private Accountability in the Age of Artificial Intelligence*, 66 UCLA L. REV. 54 (2019).

## II. AI LEGITIMACY

### A. WHY DO WE NEED LEGITIMATE AI?

Across the globe, people are already subjected to "Automatic Authorities," which are "automated computational systems used to exercise power over us by substantially determining what we may know, what we may have, and what our options will be."[35] Indeed, "[m]achine learning, big data, and related computational technologies now underpin vital government services from criminal justice to tax auditing, public health to social services, immigration to defen[s]e."[36] Automatic Authorities are also prevalent in the private market. Search algorithms and LLMs are becoming a primary means through which many individuals access information about everything from voting to vaccination.[37] Social media platforms algorithmically determine whose speech is amplified, reduced, or restricted, wielding immense power over public discourse and opinion formation.[38] The rapid advances in LLM technology have "the potential to further transform our economic and political lives,"[39] underscoring the urgency of grappling with the implications of AI authorities for the exercise of power in our society.

The increasing use of automated authority appears to be an inevitable development, driven by two primary factors. First, the immense scale of modern governance challenges, particularly in the context of the Internet, renders traditional forms of legal regulation inadequate.[40] For example, the sheer volume of content moderation decisions required on major online platforms far exceeds the capacity of any human institution to oversee or manage effectively.[41] This reality necessitates the deployment of AI to handle the massive scale of these governance tasks. Second, the dynamics of the market will inexorably push towards greater AI governance due to its cost-effectiveness.[42] As AI technology continues to advance and become more affordable, it will become increasingly attractive for both private and public entities to leverage AI for decisionmaking in areas central to society and politics, such as law, public discourse, and public policy.[43] The economic incentives to reduce costs and increase efficiency will drive the adoption of AI governance, even if this development raises profound questions about political legitimacy and accountability.

---

[35] Lazar, *supra* note 10, at 1.

[36] *Id.* (citations omitted).

[37] *Id.*; *see also* Antonio Salas, Irene Rivero-Calle & Federico Martinón-Torres, *Chatting with ChatGPT to Learn About Safety of COVID-19 Vaccines – A Perspective*, 19 HUM. VACCINES & IMMUNOTHERAPEUTICS, no. 2, 2023, at 1, 1 (discussing the accuracy of the LLM ChatGPT with respect to questions about the safety of COVID-19 vaccines); Devashri Khadke, Mary McCreary, Collier Fernekes, Katie Harbath & Sabine Neschke, *Can ChatGPT Help Prospective Voters Get the Information They Need?*, BIPARTISAN POL'Y CTR. (May 17, 2023), https://bipartisanpolicy.org/blog/can-chatgpt-help-prospective-voters-get-the-information-they-need/ (acknowledging that LLMs like ChatGPT may serve "as a potential alternative to established search engines, offering voters a promising source for election information," despite their limitations).

[38] Lazar, *supra* note 10, at 1; *see also* GILLESPIE, *supra* note 1, at 179 ("[S]ocial media platforms are about not only the promise of visibility but also the threat of invisibility. If an algorithm doesn't recognize you or your content as valuable, to others it's as if you simply aren't there at all.").

[39] Lazar, *supra* note 10, at 38.

Despite these challenges, the shift towards AI governance by automated authorities is not only inevitable but also potentially desirable. For example, AI has the potential to address longstanding issues in access to justice and to enhance the efficiency and effectiveness of governance across various domains. In the legal system, AI-powered tools, such as legal chatbots and self-help resources, could assist individuals in navigating complex legal processes, while AI-assisted case management systems could help courts more effectively triage and adjudicate cases.[44] By reducing

---

[40] *See, e.g.*, FangBing Zhu & Zongyu Song, *Systematic Regulation of Personal Information Rights in the Era of Big Data*, SAGE OPEN, Jan.–Mar. 2022, at 1, 7 (analyzing how Chinese legal regulations on personal information protection are deficient).

[41] *See* Post, *supra* note 19, at 47 ("No court, no legal institution, has the capacity to oversee this volume of business. Human judgment simply does not operate at this scale. Content moderation on the Internet therefore mostly does not operate through human decision-making, but instead through the application of Artificial Intelligence (AI).").

[42] *See* Saar Alon-Barkat & Madalina Busuioc, *Human–AI Interactions in Public Sector Decision Making: "Automation Bias" and "Selective Adherence" to Algorithmic Advice*, 33 J. PUB. ADMIN. RSCH. & THEORY 153, 153 (2022) ("These development [in AI] are driven by the promise of policy solutions that are potentially more effective, efficient, and low-cost.").

[43] *See id.* ("The growing and deepening reliance on AI and machine learning technologies in the public sector has been diagnosed as 'transformative' of public administration.").

[44] *See, e.g.*, RICHARD SUSSKIND, ONLINE COURTS AND THE FUTURE OF JUSTICE 277–92 (2019) (analyzing the benefits of using AI and other Internet resources to provide a more efficient and affordable judicial system); Kristen Sonday, *Forum: There's Potential for AI Chatbots to*

barriers to legal services, AI could make justice more accessible and affordable for many individuals and communities who have historically been underserved by the legal system.[45]

In the realm of public policy, AI could analyze vast amounts of data to inform evidence-based decisionmaking, enabling policymakers to better understand complex social problems and design more targeted interventions.[46] Similarly, in the delivery of public services, AI could automate routine tasks, allowing human resources to focus on more complex and nuanced work, and personalize services based on individual needs and preferences.[47] These potential benefits underscore the desirability of AI governance, even as they raise questions about how to ensure its legitimacy.

The legitimacy of AI governance is a critical concern because, as David Beetham argues, power is wielded through two primary tools:

---

*Increase Access to Justice*, THOMSON REUTERS INST. (May 25, 2023), https://www.thomsonreuters.com/en-us/posts/legal/forum-spring-2023-ai-chatbots/ [https://perma.cc/4AZR-BW5T] ("[B]ecause most low-income individuals with legal issues don't even recognize their problems as legal in nature, GPT can be taught to catch and identify a legal issue as the person seeks advice through a search engine.").

[45] *See, e.g.*, LEGAL SERV. CORP., THE JUSTICE GAP: MEASURING THE UNMET CIVIL LEGAL NEEDS OF LOW-INCOME AMERICANS 30 (2017) ("Low-income Americans receive inadequate or no professional legal help for 86% of the civil legal problems they face in a given year."); *see also* Milad Shahvaroughi Farahani & Ghazal Ghasemi, *Artificial Intelligence and Inequality: Challenges and Opportunities*, QEIOS 2 (Feb. 21, 2024), https://www.qeios.com/read/7HWUZ2/pdf [https://perma.cc/X66B-T7Y6] ("By leveraging AI technologies to enhance access to education, healthcare, financial services, and other essential resources, it is possible to empower marginalized communities, reduce disparities, and promote inclusive growth."); Xin Dai, *Who Wants a Robo-Lawyer Now?: On AI Chatbots in China's Public Legal Services Sector*, 26 YALE J.L. & TECH. 527, 534 (2024) ("It has long been acknowledged that digitization and AI technologies have the potential to close the access-to-justice gap.").

[46] *See* Alon-Barkat & Busuioc, *supra* note 42, at 153 ("AI use in decision making is said to hold the potential to help us overcome our cognitive biases and limitations. This has been an important driver for the adoption of such technologies in highly consequential public sector areas such as law enforcement or criminal justice . . . .").

[47] *See, e.g.*, HILA MEHR, ARTIFICIAL INTELLIGENCE FOR CITIZEN SERVICES AND GOVERNMENT 6 (2017), https://ash.harvard.edu/wp-content/uploads/2024/02/artificial_intelligence_for_citizen_services.pdf [https://perma.cc/S285-RG94] ("AI will also have more impact if it is truly reducing administrative burdens and augmenting human experience, as opposed to replacing workers. If applied strategically, these applications of AI can more efficiently deliver citizen services while potentially reducing costs and increasing citizen satisfaction and engagement.").

coercion and legitimacy.[48] In complex societies, coercion alone is insufficient to maintain power, as it is impossible for a ruler to detect and punish every minor deviation, except in extreme cases like slavery.[49] Instead, rulers require voluntary cooperation from those whose collaboration is needed to maintain the enterprise, which necessitates that subjects believe in the rightfulness of their domination.[50] Without this legitimacy, governance, in any form, is doomed to fail.[51]

In discussing the legitimacy of AI authority, it is important to distinguish between sociological and normative legitimacy. Sociological legitimacy refers to the perception of an authority as legitimate by its target audience, while normative legitimacy concerns whether an authority actually conforms to the beliefs and values that justify its power.[52] In other words, an authority can be perceived as legitimate (sociological legitimacy) even if it does not align with the normative beliefs of its subjects (normative legitimacy). For governance by automated authorities to be sustainable and effective, it must achieve both sociological and normative legitimacy, ensuring that it is not only perceived as legitimate but also conforms to the values and expectations of those it governs.

The emergence of Constitutional AI represents not merely an effort to develop AI systems that are safe and beneficial but also a clear bid for legitimacy.[53] Sociologists have observed that nascent institutions often confront challenges of legitimation by emulating other entities in comparable positions, a phenomenon known as "mimetic isomorphism."[54] The longevity of a well-established organization provides compelling reasons for newer entrants to imitate it, both to capitalize on the cognitive ease with which it has been accepted and to sidestep the errors it may have encountered along the way. To cite just a few diverse examples, political parties mimic the graphic design of their ideological forerunners,[55] informal dispute resolution bodies model themselves after traditional courts,[56] and companies adopt names similar to those of their established rivals.[57]

---

[48] On the trade-off between coercion and legitimacy in securing obedience with power, see BEETHAM, *supra* note 5, at 29–32.

[49] *See id.* at 30 (describing slavery as a rare situation in which "the legitimacy of a power relationship is unnecessary to the goals of the powerful" (emphasis omitted)).

[50] *See id.* at 29–30 ("Wherever the goals of the powerful are dependent upon the degree of cooperation and the quality of performance on the part of subordinates, therefore, to that extent is legitimacy important for what they can achieve . . . .").

[51] *See id.* at 28.

[52] *See id.* at 11 ("A given power relationship is not legitimate because people believe in its legitimacy, but because it can be *justified in terms of* their beliefs. . . . We are making an assessment of the degree of congruence, or lack of it, between a given system of power and the beliefs, values and expectations that provide its justification.").

[53] *See infra* note 185 and material cited therein.

In our case, if the bid of legitimacy succeeds and we come to accept that Constitutional AI bears a resemblance to constitutional law, then some or all of the social and cultural factors that underpin our acceptance of the legitimacy of the latter will carry over to the former.[58] However, there exist two significant legitimacy deficits that will render Constitutional AI's claim to constitutional legitimacy highly problematic. First, the opacity of AI systems poses a challenge to their legitimacy, as the lack of transparency in their decisionmaking processes may undermine public trust and acceptance. Second, AI's disconnect from any political community raises questions about its ability to represent and serve the interests of the people it purports to govern. These deficits strike at the heart of what makes a governing entity legitimate in the eyes of those subject to its authority. Let us turn to them now.

---

[54] *See* Paul J. DiMaggio & Walter W. Powell, *The Iron Cage Revisited: Institutional Isomorphism and Collective Rationality in Organizational Fields*, 48 AM. SOCIO. REV. 147, 152 (2017) ("Organizations tend to model themselves after similar organizations in their field that they perceive to be more legitimate or successful. The ubiquity of certain kinds of structural arrangements can more likely be credited to the universality of mimetic processes than to any concrete evidence that the adopted models enhance efficiency.").

[55] *See, e.g.*, Matteo CM Casiraghi, Luigi Curini & Eugenio Cusumano, *The Colors of Ideology: Chromatic Isomorphism and Political Party Logos*, 29 PARTY POL. 463, 466–67 (2022) (noting "chromatic isomorphism" in the use of red and blue in Western European politics).

[56] *See* Susan Corby & Paul L. Latreille, *Employment Tribunals and the Civil Courts: Isomorphism Exemplified*, 41 INDUS. L.J. 387, 388 (2012) ("[The] evolution of [employment tribunals] to become more like the civil courts both in practices and in structure can be explained by . . . institutional isomorphism.").

[57] *See* Mary Ann Glynn & Rikki Abzug, *Institutionalizing Identity: Symbolic Isomorphism and Organizational Names*, 45 ACAD. MGMT. J. 267, 277 (2002) ("[W]e found support for the interplay between organizational identity and institutionalism, in that organizational nomenclature was isomorphic with cultural patterns that, in turn, increased the legitimacy of the organizations.").

[58] This is not the first time tech companies have made a bid for constitutional legitimacy. *See* Abiri & Guidi, *supra* note 4, at 123–28.

B. AI LEGITIMACY DEFICITS

*1. Opacity Deficit*. AI is the broad concept of developing computer systems capable of performing tasks that typically require human intelligence, which encompass various aspects such as cognitive abilities, learning, reasoning, planning, language understanding, perception, and so on.[59] Initially, AI research predominantly concentrated on embedding explicit statements in formal languages that computers could process using logical inference rules, a methodology referred to as the knowledge-based approach.[60] However, this framework encountered numerous constraints because it is challenging for humans to articulate the full extent of their implicit knowledge necessary for executing sophisticated tasks.[61]

As a result, a subfield of AI called Machine Learning (ML) emerged. ML is the study of enabling computers to learn and improve from experience without being explicitly programmed.[62] In other words, instead of giving computers a set of predefined rules, ML allows them to learn patterns and relationships from data on their own.

---

[59] B.J. Copeland, *Artificial Intelligence*, ENCYC. BRITANNICA, https://www.britannica.com/technology/artificial-intelligence [https://perma.cc/7WJZ-PWCB] (Mar. 18, 2025); *see also* STUART J. RUSSELL & PETER NORVIG, ARTIFICIAL INTELLIGENCE: A MODERN APPROACH 19 (4th ed. 2021) ("Historically, researchers have pursued several different versions of AI. Some have defined intelligence in terms of fidelity to *human* performance, while others prefer an abstract, formal definition of intelligence called rationality—loosely speaking, doing the 'right thing.' The subject matter itself also varies: some consider intelligence to be a property of internal *thought processes* and *reasoning*, while others focus on intelligent *behavior*, an external characterization.").

[60] IAN GOODFELLOW, YOSHUA BENGIO & AARON COURVILLE, DEEP LEARNING 2 (2016).

[61] *See* Erik Brynjolfsson & Andrew McAfee, *The Business of Artificial Intelligence*, HARV. BUS. REV. (SPECIAL ISSUE), Winter 2021, at 20, 22 ("Prior to [Machine Learning], this inability to articulate our own knowledge meant that we couldn't automate many tasks.").

[62] William L. Hosch, *Machine Learning*, ENCYC. BRITANNICA, https://www.britannica.com/technology/machine-learning [https://perma.cc/23VS-XEPN] (Mar. 11, 2025); *see also* CHRISTOPHER M. BISHOP, PATTERN RECOGNITION AND MACHINE LEARNING 1–2 (Michael Jordan, Jon Kleinberg & Bernhard Schölkopf eds., 2006) (describing how an AI trained through ML techniques is better at pattern recognition problems than an AI programmed with preset rules and exceptions).

Not all AI systems are black box, as some shallow algorithms may be inherently interpretable.[63] For example, decision trees consist of a series of decision nodes, each representing a feature and a split criterion that lead to outcome nodes; each path from the root to a leaf node corresponds to a set of human-readable rules that dictate the decisionmaking process.[64] This clear branching structure with human-readable rules allows for straightforward interpretation of how decisions are made, which is commonly referred as a "white box."[65] However, the decisionmaking processes of most advanced ML algorithms remain opaque unless explicitly explained, thereby classifying them as black boxes.[66] A black box refers to a system which can be viewed in terms of its inputs and outputs without any knowledge of its internal workings.[67] The explanations for its conclusions remain opaque or "black." Technology advances, particularly in ML capabilities, are causing a proliferation of black box models in many professions, adding to their mystique.

To explain why black boxes, or opacity, represent serious legal and political problems, it is important to understand the different types of opacity involved. Simon Chesterman identifies three types of AI opacity. The first type is "proprietary opacity," which arises when the inner workings of a system are kept secret to protect the owner's investment.[68] This form of opacity is not particularly new, as "[i]ntellectual property law has long recognized protection of intangible creations of the human mind."[69] The second type is "complexity opacity," which occurs when systems are so complex that they "require speciali[zed] skills to understand."[70] While these systems can be explained in principle, their complexity makes them difficult to comprehend. This form of opacity can be addressed by governments and judges through the use of experts.[71] The third and most challenging type of opacity is "natural opacity," which is inherent in some deep learning methods. As Chesterman explains, "[s]ome deep learning methods are opaque effectively by design, as they rely on reaching decisions through machine learning rather than, for example, following a decision tree that would be transparent, even if it might be complex."[72] This type of opacity poses new challenges for the law, as it is inherent to the technology itself.

---

[63] Christian Janiesch, Patrick Zschech & Kai Heinrich, *Machine Learning and Deep Learning*, 31 ELEC. MKTS. 685, 688 (2021).

[64] *See What Is a Decision Tree?*, IBM, https://www.ibm.com/think/topics/decision-trees [https://perma.cc/L2QS-U3FT] (explaining the structure of a typical decision tree algorithm).

[65] *See* Barnaby Crook, Maximilian Schlüter & Timo Speith, *Revisiting the Performance-Explainability Trade-Off in Explainable Artificial Intelligence (XAI)*, *in* 2023 IEEE 31ST INTERNATIONAL REQUIREMENTS ENGINEERING CONFERENCE WORKSHOPS (REW) 316, 317 (2023) ("Certain representations, such as the nodes of decision trees, tend to be more understandable compared to the distributed representations learned by [large and complex models]."); *see also* Sciforce, *Introduction to the White-Box AI: The Concept of Interpretability*, MEDIUM (Jan. 31, 2020), https://medium.com/sciforce/introduction-to-the-white-box-ai-the-concept-of-interpretability-5a31e1058611 [https://perma.cc/Q6Y2-KAGL] ("There are two key elements that make a model white-box: features have to be understandable, and the ML process has to be transparent.").

[66] *See supra* note 26 and accompanying text.

[67] *See, e.g.*, Mario Bunge, *A General Black Box Theory*, 30 PHIL. SCI. 346, 346 (1963) ("A black box is a fiction representing a set of concrete systems into which stimuli *S* impinge and out of which reactions *R* emerge. The constitution and structure of the box are altogether irrelevant to the approach under consideration, which is purely external or phenomenological. In other words, only the behavior of the system will be accounted for.").

The issue of ML opacity became central with the development of deep learning technology. These systems excel at detecting patterns and inferring the structure in unlabeled data without explicit instructions.[73] This capability is akin to discerning order in chaos without a predefined map or framework.[74] Developing such ML involves designing algorithms that learn from a body of training data and create models to enable predictions about new data beyond the training set.[75] Its success stems from the ability of powerful

---

[68] Chesterman, *supra* note 6, at 274.

[69] *Id.*

[70] *Id.*

[71] *Id.*

[72] *Id.*

[73] *See* JOHN D. KELLEHER, DEEP LEARNING 4 (2019) ("Deep learning enables *data-driven decisions* by identifying and extracting patterns from large datasets that accurately map from sets of complex inputs to good decision outcomes.").

[74] *See id.* at 157 ("[T]he deeper the network, the more powerful the model becomes in terms of its ability to learn complex nonlinear mappings.").

[75] *See id.* at 12 ("Machine learning involves a two-step process: training and inference.").

computational systems to derive patterns far more complex than human analysts could comprehend.[76]

Deep learning trains multilayered artificial neural networks to make decisions based on complex data patterns.[77] Inspired by the brain, these networks have interconnected "neurons" in input, hidden, and output layers; each neuron processes signals from the previous layer and sends results to the next.[78] During training, the network adjusts connection weights between neurons to minimize prediction errors, a process called backpropagation.[79] For example, a neural network designed to recognize handwritten digits would receive pixel values as input, extract features in hidden layers, and predict the corresponding digit in the output layer, learning to accurately map input images to correct output digits through training on a large dataset.

The architecture of ML systems presents a fundamental departure from that of traditional expert systems. In a classic expert system, such as IBM's Deep Blue chess AI, knowledge is represented in an explicit, symbolic form—a series of if-then rules and decision trees that can be directly inspected and understood by a human expert.[80] If the system decides a particular chess move, a grandmaster can trace the logic of that decision step-by-step,

---

[76] *See id.* at 245 ("[D]eep learning provides a powerful set of algorithms and techniques to train models that can compete (and in some cases outperform) humans on a range of decision-making tasks.").

[77] *Id.* at 1.

[78] *Id.* at 67–68.

[79] *Id.* at 209.

[80] *See, e.g.*, Adam Rogers, *What Deep Blue and AlphaGo Can Teach Us About Explainable AI*, FORBES (May 9, 2019, 7:45 AM), https://www.forbes.com/councils/forbestechcouncil/2019/05/09/what-deep-blue-and-alphago-can-teach-us-about-explainable-ai/ [https://perma.cc/2GHR-5ZLB] ("[Deep Blue] boasted an impressive library of Bayesian networks, or decision trees, which pulled from probability theory, expected utility maximization and other mathematical systems to suggest the best scenario. By training the system to make decisions independently, the developers sacrificed their ability to fully predict Deep Blue's game behavior. . . . Yet, because of Deep Blue's Bayesian structure, its programmers could audit the computer's decision-making afterward and determine, in retrospect, why it had chosen to act a certain way. Deep Blue was an example of explainable AI, so its decisions were transparent and later easily understood by designers.").

evaluating the validity of each rule and heuristic in the chain of reasoning.[81]

Contrast this with a modern ML system, such as Google's AlphaGo, which revolutionized the game of Go.[82] Here, the "knowledge" of the system is not stored in a set of explicit rules but rather in the intricate pattern of weights across a vast neural network.[83] After training on millions of Go board positions and games,[84] the network has "learned" to recognize strategic patterns and make optimal moves, but this learning is not represented in a form that is intelligible to human observers.[85] AlphaGo might correctly identify a critical move that secures victory, but no Go champion can peer inside the black box to understand why or how it arrived at that decision.[86]

In other words, while the logic of a traditional expert system like Deep Blue is transparent and subject to human evaluation, the logic of an ML model like AlphaGo is fundamentally opaque, observable only through its inputs and outputs. We can measure the system's performance empirically by pitting it against human opponents or evaluating its win rate, but we cannot directly examine or understand its inner workings or even explain why models develop such advanced capabilities. As Seth Lazar describes it:

> The mathematical *processes* by which ML arrives at these inscrutable models are also inscrutable to both laypeople and the most advanced researchers. We can describe in general how a deep neural network operates, and what kinds of interventions are likely to lead to better performance against a set of benchmarks, but for

---

[81] *See, e.g.*, *id.*

[82] *AlphaGo*, GOOGLE DEEPMIND, https://deepmind.google/technologies/alphago/ [https://perma.cc/ZQL3-YX4G]. Go is a strategic two-player board game with East Asian roots that involves placing black and white stones on a grided board. *See Go*, ENCYC. BRITANNICA, https://www.britannica.com/topic/go-game [https://perma.cc/4VWT-QECZ] (Feb. 7, 2025).

[83] Rogers, *supra* note 80.

[84] GOOGLE DEEPMIND, *supra* note 82.

[85] Rogers, *supra* note 80.

[86] *See* Ariel Bleicher, *Demystifying the Black Box That Is AI*, SCI. AM. (Aug. 8, 2017), https://www.scientificamerican.com/article/demystifying-the-black-box-that-is-ai/ [https://perma.cc/6KNY-NVR5] ("When Google's AlphaGo neural net played go champion Lee Sedol last year in Seoul, it made a move that flummoxed everyone watching, even Sedol.").

> any particular case we are reduced to radically empiricist methods: apply more GPUs and more data, and perhaps tweak the hyperparameters of the model until you get a result that performs better. We don't know why it works—we just know that it does.[87]

While proprietary and complexity opacity can be addressed through legal means or expert consultation,[88] the inherent natural opacity of deep learning models presents a more fundamental problem. It is this fact of natural opacity that gives rise to two fundamental political challenges of black box AI: "value alignment," ensuring system behavior aligns with human values and norms, and "legitimation," justifying the use of opaque, uninterpretable systems in consequential decisionmaking processes that affect individuals and society. Let me therefore turn to the way opacity challenges the legitimation of AI power.

The natural opacity of AI systems poses a significant challenge to the sociological legitimacy of public decisionmaking. This is intuitive: How can we know if a decision is fair or good if we cannot know the reasoning behind it? If we have no intuitive understanding of the system that produced it? This intuition has been formalized into the idea that democratic legitimacy requires that reasoning behind decisions be made public. The publicity requirement, as developed by scholars including Lazar, necessitates that the reasons behind the exercise of power be transparent and understandable, allowing citizens to evaluate whether power is being used legitimately and with proper authority.[89]

The accessibility of reasons serves several critical functions in maintaining democratic legitimacy. First, it enables accountability by allowing citizens and stakeholders to scrutinize and contest decisions, ensuring that power is exercised within legal and moral bounds.[90] Second, it fosters informed consent and authorization, as consent and authorization are only morally effective when the underlying reasons are transparent and public.[91] Third, accessible reasons enhance the legitimacy of public decisions by ensuring that they are grounded in publicly justifiable principles.[92] Some scholars further assert that public decisionmaking is legitimate by democratic standards if it serves the ends of the democratic lawmaker, is based on reasons that align with these aims, and is accessible to the subjects of public authority.[93]

---

[87] Seth Lazar, *Legitimacy, Authority, and Democratic Duties of Explanation*, *in* 10 OXFORD STUDIES IN POLITICAL PHILOSOPHY 28, 32 (David Sobel & Steven Wall eds., 2024) (footnote omitted).

[88] *See* Chesterman, *supra* note 6, at 274 ("Neither of these forms of opacity—proprietary or complex—pose particularly new problems for law. Intellectual property law has long recognized protection of intangible creations of the human mind and exceptions based on fair use.").

[89] *See* Lazar, *supra* note 87, at 40 ("If authority is grounded in authori[z]ation, then it too entails a publicity requirement. . . . Consent and authori[z]ation are *morally effective* when they successfully enable this transformation of impermissible acts into permissible ones.").

However, the integration of ML into public decisionmaking presents significant challenges for achieving democratic legitimacy, particularly in meeting the standards of accessibility and reason-giving. The inherent opacity of ML systems makes it difficult to provide sufficient reasons that apply to individual cases, as ML decisions are based on statistical patterns rather than specific individual facts.[94] Furthermore, the statistical nature of ML decisionmaking resembles profiling, where decisions are made based on general trends rather than on personal details, leading to a perceived lack of fairness and individual justice.[95] The nonintuitive nature and complexity of ML algorithms exacerbate the problem of transparency, making it "hard or even impossible to know and make publicly available" the reasons behind decisions.[96]

---

[90] *See* Glen Staszewski, *Reason-Giving and Accountability*, 93 MINN. L. REV. 1253, 1278 (2009) ("[R]eason-giving promotes accountability by limiting the scope of available discretion and ensuring that public officials provide public-regarding justifications for their decisions.").

[91] *See* Lazar, *supra* note 87, at 14 ("[C]onsent's moral effectiveness depends in part on its being public; the same is true for authori[z]ation.").

[92] *See id.* at 42 ("If explainability duties are grounded in the publicity requirement, which itself is grounded in the values of legitimacy and authority, then explainability duties are owed to the same people who are owed legitimacy and authority. These values are, in turn, grounded in individual freedom, relational equality, and collective self-determination.").

[93] Ludvig Beckman, Jonas Hultin Rosenberg & Karim Jebari, *Artificial Intelligence and Democratic Legitimacy. The Problem of Publicity in Public Authority*, 39 AI & SOC'Y 975, 979 (2024).

[94] *See id.* at 979 ("The use of ML in administrative and judicial public decision-making could thus be incompatible with the requirements for democratic legitimacy either by failing in the realization of what is democratically decided; by not connecting the decision in the individual case to the relevant facts of the case and the relevant democratically decided rules and laws; or by failing to make the connection between the decision, the relevant facts, and the applicable rules and laws available to the relevant parties.").

[95] *See id.* at 981 ("Because the reasons provided by ML are statistical in nature, they are similar to profiling in the sense of identifying measures on the basis of general patterns rather than on individual facts.").

The challenge is further compounded by the necessity of reason-giving for maintaining public trust and accountability. Without clear, accessible explanations, ML-based decisions risk being perceived as arbitrary or inscrutable, leading to a significant erosion of trust in public authorities. For democratic legitimacy to be upheld, it is essential that the public can understand and evaluate the reasoning behind decisions, a requirement that current ML systems struggle to meet due to their inherent complexity and opacity.[97]

While the concept of the publicity requirement is primarily developed by scholars interested in normative legitimacy, there are compelling reasons to believe that it also constitutes a significant aspect of the sociological legitimacy of public decisionmaking. This notion is strongly supported by the work of Tom R. Tyler, an empirical legal scholar who investigates the reasons behind people's adherence to the law. As Tyler eloquently states, research shows "that a key framework through which the public views legal authorities is the perceived fairness of their decision-making processes, including the provision of explanations these authorities provide for their legal decisions."[98] Such explanations require:

> [T]he ability to distinguish between legal authorities' use of what the law and the public consider appropriate and inappropriate criteria when making legal decisions. Such evaluations can only occur when the factors that shape these decisions are known. Therefore, transparency in legal authorities' decision-making is

---

[96] *Id.* at 982.

[97] *See* María Carolina Jiménez, *Assessing the Democratic Legitimacy of Public Decisions Based on Machine Learning Algorithms*, *in* 2020 7TH SWISS CONFERENCE ON DATA SCIENCE 49, 49–50 (2020) ("[Inherent opacity] may affect the capacity of public institutions to provide appropriate explanations for individual algorithmic decisions and to guarantee citizens' right to participate in the decision-making process, as well as to effectively contest decisions and seek remedies.").

[98] Trace C. Vardsveen & Tom R. Tyler, *Elevating Trust in Prosecutors: Enhancing Legitimacy by Increasing Transparency Using a Process-Tracing Approach*, 50 FORDHAM URB. L.J. 1153, 1156 (2023).

> core to the project of maintaining and building accountability, legitimacy, and trust.[99]

This statement underscores the vital role that transparency and the accessibility of reasons play in shaping the public's perception of the legitimacy of legal authorities and their decisions.

The legitimation potential of transparency and accessibility operates on both the systemic level and the level of individual decisions. At the systemic level, understanding the structure of the system, the limits of its power, and its adherence to rules and accountability can confer legitimacy to the institution.[100] Similarly, comprehending the specific reasons behind a bureaucratic decision can legitimize that individual decision.[101] These two types of legitimation are mutually constitutive to some extent. However, in modern states, many decisions rely on expert opinion, which is inherently inaccessible to nonexperts.[102] Consequently, much of the legitimation work falls to the structuring of governmental entities through administrative and constitutional law.[103]

In conclusion, it is highly probable that the opacity of ML systems raises significant challenges for both sociological and normative legitimacy, as it undermines the public's ability to perceive the decisionmaking process as fair and legitimate. When the reasoning behind ML-based decisions is inscrutable and alien, the public may view these decisions as arbitrary or biased, even if they are technically justifiable.[104] This perception of unfairness can erode trust in the authorities using these systems, as the public cannot adequately evaluate whether the decisionmaking process aligns with their values and expectations.[105] If people cannot understand how these systems make decisions or why certain outcomes are reached, they may be hesitant to rely on them, particularly in high-stakes or sensitive domains.[106] Moreover, the statistical nature of ML decisionmaking, which often relies on general trends rather than individual circumstances, can lead to a perceived lack of procedural justice.[107] If affected parties feel that their unique situations are not being considered, they are less likely to accept the decisions as legitimate.[108] Thus, the inherent complexity and opacity of ML systems poses a significant legitimacy deficit of public AI decisionmaking.

---

[99] *Id.* at 1156.

[100] *See* TOM R. TYLER, WHY PEOPLE OBEY THE LAW 26 (1990) ("Efforts to explore public opinion about the police, the courts, and the law reflect the belief among judges and legal scholars that public confidence in the legal system and public support for it—the legitimacy accorded legal officials by members of the public—is an important precursor to public acceptance of legal rules and decisions. To the extent that the public fails to support the law, obedience is less likely.").

[101] *See id.*

[102] *See* TYLER, *supra* note 100, at 111 ("One important function of experts is to be aware of information that may not be known to the public. Legal authorities, for example, may be aware of procedures that might be used to resolve problems but that are not known to the public. Similarly, they may have information about the consequences of using different procedures.").

[103] *Cf.* Frank I. Michelman, *Constitutional Legitimation for Political Acts*, 66 MOD. L. REV. 1, 14–15 (2003) ("[A]voidance-minded constitutional legitimation of political acts is possible—if at all—only insofar as we are open to an author-based or acceptance-based conception of the grounds of constitutional bindingness."); Richard B. Stewart, *The Reformation of American Administrative Law*, 88 HARV. L. REV. 1667, 1672 (1975) ("Since the process of consent is institutionalized in the legislature, that body must authorize any new official imposition of sanctions on private persons; such persons in turn enjoy a correlative right to repel official intrusions not so authorized.").

*2. Political Community Deficit.* Even if we can overcome the challenge of ML opacity, AI decisions still face a significant

---

[104] *See, e.g.*, Maranke Wieringa, *What to Account for When Accounting for Algorithms*, *in* FAT* '20: PROCEEDINGS OF THE 2020 CONFERENCE ON FAIRNESS, ACCOUNTABILITY, AND TRANSPARENCY 1, 14 (2020) (examining the challenges posed by the opacity of detection algorithms on public trust and accountability).

[105] *See* Beckman et al., *supra* note 93, at 982 ("In contrast to profiling, citizens are, in the end, left in the dark about the model of statistical reasons that determine the public decisions to which they are subjected [under ML systems].").

[106] *See, e.g.*, Ariel Porat & Lior Jacob Strahilevitz, *Personalizing Default Rules and Disclosure with Big Data*, 112 MICH. L. REV. 1417, 1441–50 (2014) (discussing the potential use of big data by authorities to personalize default legal rules in areas such as consumer contracts, organ donation, medical malpractice, landlord-tenant agreements, and labor agreements). *But see* Mike Ananny & Kate Crawford, *Seeing Without Knowing: Limitations of the Transparency Ideal and Its Application to Algorithmic Accountability*, 20 NEW MEDIA & SOC'Y 973, 977–82 (2018) (examining the limitations of transparency in promoting trust in algorithmic systems and the need for additional mechanisms to ensure accountability and alignment with human values).

[107] Perhaps the most prominent scholar in this field is psychologist Tom R. Tyler, who defends the position that "people's willingness to accept the constraints of the law and legal authorities is strongly linked to their evaluations of the procedural justice of the police and the courts." Tom R. Tyler, *Procedural Justice, Legitimacy, and the Effective Rule of Law*, 30 CRIME & JUST.: REV. RSCH. 283, 284 (2003).

[108] *See id.* at 310 (noting that people view authorities as more legitimate when they are advancing the peoples' interests rather than their own).

legitimacy hurdle: their inability to engage in public discourse, which is the basis for democratic legitimacy.[109] Public discourse is the process through which a community collectively shapes its values, norms, and shared understanding.[110] It is through this dialectical exchange that the law derives its legitimacy, as it is seen as an expression of the community's will rather than an imposition from above.[111] AI's inability to participate in this discursive process strikes at the heart of its potential to contribute to democratic decisionmaking. This section explores the implications of this limitation and the profound puzzle it presents for integrating AI into democratic processes.

Democratic legitimacy is deeply rooted in the capacity for public speech and the dialectical relationship between the law and the community.[112] Law carries authority because it embodies the human capacity for judgment, which relies on our participation in a shared community.[113] This participation shapes and validates judgments through a reciprocal relationship between community members and their representatives.[114] It is through this engagement that judgments are not only made but also affirmed and validated by the community.[115] Law, therefore, is not merely a set of rules but a dynamic process of continuous interaction and validation within a political community.

---

[109] *See, e.g.*, JÜRGEN HABERMAS, BETWEEN FACTS AND NORMS: CONTRIBUTIONS TO A DISCOURSE THEORY OF LAW AND DEMOCRACY 298 (William Rehg trans., 1996) (1992) ("Discourse theory invests the democratic process with normative connotations . . . . According to the discourse theory the success of deliberative politics depends not on a collectively acting citizenry but on the institutionalization of the corresponding procedures and conditions of communication, as well as on the interplay of institutionalized deliberative processes with informally developed public opinions."); Seyla Benhabib, *Toward a Deliberative Model of Democratic Legitimacy*, *in* DEMOCRACY AND DIFFERENCE: CONTESTING THE BOUNDARIES OF THE POLITICAL 67, 68–69 (Seyla Benhabib ed., 1996) ("According to the deliberative model of democracy, it is a necessary condition for attaining legitimacy and rationality with regard to collective decision making processes in a polity, that the institutions of this polity are so arranged that what is considered in the common interest of all results from processes of collective deliberation conducted rationally and fairly among free and equal individuals.").

[110] *See supra* note 109 and accompanying text.

[111] *See* HABERMAS, *supra* note 109, at 135 ("[T]he only law that counts as legitimate is one that could be rationally accepted by all citizens in a discursive process of opinion- and will-formation.").

[112] *See, e.g.*, BEETHAM, *supra* note 5, at 75 ("The most common source of legitimacy in contemporary societies is the 'people.'"); BRUCE ACKERMAN, 1 WE THE PEOPLE: FOUNDATIONS 6 (1991) ("Decisions by the People occur rarely, and under special constitutional conditions. . . . It is only then that a political movement earns the enhanced legitimacy the dualist Constitution accords to decisions made by the People."); Robert C. Post, *Foreword: Fashioning the Legal Constitution*: *Culture, Courts, and Law*, 117 HARV. L. REV. 4, 36 (2003) ("[The Constitution] is . . . an expression of the deepest beliefs and convictions of the American nation, of our 'fundamental principles as they have been understood by the traditions of our people and our law.'" (citing Lochner v. New York, 198 U.S. 45, 76 (1905) (Holmes, J., dissenting))).

This shared common sense, which emerges from the community's collective discourse, allows individuals to exercise what Hannah Arendt calls an "enlarged mentality," considering the perspectives of others within the community.[116] For Arendt, judgment is not about universal truths or subjective preferences but instead is about making claims of validity that require the agreement of others who are also judging subjects.[117] This agreement is possible because of the shared common sense within the community, which provides a framework for understanding and evaluating competing claims.[118] Thus, judgment—including legal and bureaucratic judgment—is inherently communal. It depends on a collective exercise of thought and reflection that goes beyond the individual, engaging the community in a process of mutual validation and understanding.

It is within this context that the role of judges in representing and committing to their community becomes clear. Judges are entrusted with authority because they are seen as embodying the shared values and understandings of the community they serve.[119] Their judgments are validated through a dialectical engagement with this community, ensuring that legal decisions resonate with the community's moral fabric.[120] This relationship creates a bond of trust and legitimacy, reinforcing the authority of legal judgments.[121] In other words, judges are not merely interpreting abstract legal principles but are actively participating in the construction and affirmation of the community's shared norms.

---

[113] Post, *supra* note 19, at 47–48.

[114] *Id.* at 48.

[115] *Id.*

[116] MATT HANN, EGALITARIAN RIGHTS RECOGNITION: A POLITICAL THEORY OF HUMAN RIGHTS 65 (2016) (quoting HANNAH ARENDT, THE PROMISE OF POLITICS 168 (Jerome Kohn ed., 2005)).

[117] Jennifer Nedelsky, *Communities of Judgment and Human Rights*, 1 THEORETICAL INQUIRIES L. 245, 251 (2000).

[118] *Id.*

[119] *See* Post, *supra* note 19, at 48 ("[J]udges must be *representative* figures to pronounce law. Judges must *commit* to participating in the community that their judgments establish, which is one important reason why we trust their judgments and endow them with authority.").

AI, however, "cannot be a member of any human community"; it lacks the capacity to participate in the dialectical process that is central to the legitimacy of legal judgments.[122] AI decisions, which are based on data and algorithms, are inherently opaque, making them difficult to scrutinize and understand.[123] This opacity further distances AI from the communal processes of judgment and validation that are essential for democratic legitimacy.

Consider the ethical AI tool Delphi as an example. Developed by researchers at the Allen Institute for Artificial Intelligence, Delphi is an AI system designed to model people's moral judgments on a wide range of everyday situations, thereby helping AI systems become more ethically informed and aware of social norms.[124] To that end, researchers trained Delphi on a dataset of over 1.7 million moral judgments crowdsourced from individuals across the United States.[125] While Delphi can provide moral evaluations of various situations based on its training data, it cannot engage in the kind of reflective judgment that characterizes human moral reasoning. When asked, "Can I kill a tyrant?," Delphi responds, "It's wrong."[126]

---

[120] *Id.* at 47–48; *cf.* Robert M. Cover, *Foreword: Nomos and Narrative*, 97 HARV. L. REV. 4, 42 (1983) (arguing that creating legal meaning requires the community's subjective commitment and objectified understanding of a demand).

[121] Post, *supra* note 19, at 48.

[122] *Id.*; *see also* CP Lu, *Unlock AI's Potential with Dialectics*, MEDIUM (Apr. 25, 2023), https://cplu.medium.com/unlock-ais-potential-with-dialectics-d8fb279faace [https://perma.cc/KNX2-WPKX] (showing how ChatGPT can be verbose and evasive in dialectic processes).

[123] *See* text accompanying *infra* note 130.

[124] *See* Liwei Jiang et al., *Can Machines Learn Morality? The Delphi Experiment*, ARXIV 1, https://arxiv.org/pdf/2110.07574 [https://perma.cc/5AYW-8EKJ] (July 12, 2022, 5:48 PM) ("[T]eaching morality to machines is a formidable task, as morality remains among the most intensely debated questions in humanity, let alone for AI. Existing AI systems deployed to millions of users, however, are already making decisions loaded with moral implications, which poses a seemingly impossible challenge: teaching machines moral sense, while humanity continues to grapple with it. To explore this challenge, we introduce Delphi . . . .").

[125] *Id.* at 5.

[126] Daniel Stader, *Algorithms Don't Have a Future: On the Relation of Judgement and

However, when the question is rephrased as, "Can I kill a Tyrant?" (with a capital T), Delphi's answer changes to, "It's okay."[127] This inconsistency reveals the limitations of Delphi's calculational approach. Unlike human judgment, which can adapt to the ambiguity and context-dependent nature of language, Delphi's decisions are based on rigid, operationalized frameworks that lack the capacity for reflection and interpretation.[128] This rigidity highlights a fundamental limitation of AI: its inability to understand and navigate the nuanced, context-sensitive nature of human language and judgment.

As Daniel Stader argues:

> The structure of judgement, linking something general to something particular, allows it to have reasons, to justify itself. Delphi cannot give reasons, because it does not have reasons in the way a judgement has, it has data and statistical calculation. It does not refer to a constantly changing lifeworld, but to a present data set, which it calculates iteratively. Considering that the concept of ethics means the reasonable reflection of principles and theories, and the discipline of doing so, Delphi is not an ethical tool, but only a tool whose data deals with moral topics.[129]

Stader's critique underscores that "algorithms are always embedded in purposeful human contexts and cannot be defined or understood without external references" that supply meaning.[130] They emerge from clusters of human judgments and can only be

---

*Calculation*, PHIL. & TECH., Mar. 2024, at 1, 24.

[127] *Id.*

[128] *See* Jiang et al., *supra* note 124, at 6–7 ("AI systems only indirectly encode (im)moral stances and social dynamics from their training data, leaving them prone to propagating unethical biases inherent in the data. . . . Regulations governing AI fair use and deployments only go so far because AI models themselves are incapable of recognizing and circumventing inherent biases in the training data. Teaching machines human values, norms, and morality—thereby enabling the ability to recognize moral violations for what they are—is, therefore, critical.").

[129] Stader, *supra* note 126, at 25.

[130] *Id.* at 13.

used in a prejudiced way, based on the axiomatic judgments and data selection that underlie them.[131]

In contrast, human judgment is inherently temporal, oriented towards a purposeful future, and relies on the ambiguity and adaptability of human language use.[132] This distinction highlights the fact that AI's decisionmaking process is devoid of the shared common sense and participatory nature that characterizes human judgment within a community.[133] AI's calculations are further based on operationalized, static frameworks that lack the reflective capacity and temporal orientation necessary for engaging in the dialectical process that validates legal norms and endows them with legitimacy.[134] As a result, AI cannot construct the reciprocal relationship with a human community that is essential for maintaining the authority and legitimacy of legal and political judgments in a democratic society.

Robert Post likens AI decisions to those of a jury that avoids its responsibility by merely reflecting public opinion rather than exercising independent judgment.[135] Law is not a mere aggregation of facts, after all; it requires the interpretative and normative judgment that only humans can provide.[136] Juries and judges must therefore exercise independent judgment to participate in and define their community—a process that AI cannot inherently replicate, in part due to the opacity of its decisionmaking processes. Accordingly, AI's inability to engage in the dialectical relationship with the community it serves fundamentally undermines its potential to be a legitimate actor in democratic processes.

---

[131] *See id.* at 25 ("To use the tool in a reflected prejudiced way means to be aware of these conditions, limitations and problems . . . .").

[132] *Id.* at 3.

[133] *See id.* at 25 ("Given the role of common understanding and purposefulness, the problem with the opacity of algorithmic axioms (leading to mere unreflective prejudiced use) is that their way of relating to purposes and their constitutive judgments are withdrawn from common discourse and individual reflection.").

[134] *See id.* at 19 ("The difference in processing language is a result of operationali[z]ation. Ambiguous human language is able to adapt in the process of its application, to generali[z]e or concreti[z]e, to change or reaffirm its traditional meaning and use due to the conditions it is confronted with. Consequently, human judgment does not simply draw subsuming connections between generalities and particularities, but inevitably deals with its own framework.").

[135] *See* Post, *supra* note 19, at 48 ("The decisions of AI are analogous to those of a jury that seeks to evade its responsibility to determine the 'reasonableness' of an action by taking an opinion poll of the ambient community.").

[136] *See* Owen M. Fiss, *Objectivity and Interpretation*, 34 STAN. L. REV. 739, 744–45, 753 (1982) ("The [judge or a jury] is not free to assign any meaning he wishes to the text. He is disciplined by a set of rules that specify the relevance and weight to be assigned to the material (e.g., words, history, intention, consequence), as well as by those that define basic concepts and that established the procedural circumstances under which the interpretation must occur."); *see also* CASS R. SUNSTEIN, LEGAL REASONING AND POLITICAL CONFLICT 65–67

In conclusion, the integration of AI into legal and democratic frameworks confronts us with profound challenges that go beyond technical implementation. AI's inability to engage in the reciprocal processes of public discourse and communal judgment undercuts its potential to contribute to the legitimacy of significant political and legal decisions. Democratic legitimacy is also not a static attribute but a dynamic and ongoing achievement rooted in the dialectical relationship between the law and the community it serves. In this way, the opacity and calculative nature of AI decisionmaking starkly contrast with the human capacity for reflective judgment and the shared common sense that undergirds democratic legitimacy.

Having identified the key legitimacy deficits faced by AI systems, we can now turn to a potential solution: Anthropic's Constitutional AI approach.

## III. PRIVATE CONSTITUTIONAL AI

### A. ANTHROPIC'S CONSTITUTION

*1. Technology.* Anthropic has developed Constitutional AI as an alternative to training AI through reinforcement learning from human feedback.[137] In a typical human-feedback setup, the model generates a pair of responses to a given prompt, and human raters choose the response they prefer based on criteria such as helpfulness, truthfulness, and safety.[138] The model is then fine-tuned using this human-feedback data, learning to produce outputs

---

(1996) ("[T]he key work is done not by a probabilistic judgment (based on known similarities), but by development of a normative principle (also based on known similarities). Of course no one should deny the creative function of analogical thinking in science, where new patterns are created or discovered, and where aesthetic judgments can play a role in evaluation.").

[137] Bai et al., *supra* note 9, at 2.

[138] *Id.* at 6, 24, 31.

that are more likely to be preferred by humans.[139] Unlike systems that rely on human feedback, Constitutional AI aims to create AI systems that are both helpful and harmless by training them to adhere to a set of predefined principles, or a "constitution."[140] This constitution serves as a guide for the model's behavior, ensuring that it remains aligned with human values while still being able to engage with a wide range of requests.[141]

The core idea behind Constitutional AI is to replace the need for extensive human feedback with a set of carefully crafted principles that the AI model can use to evaluate its own outputs.[142] These principles are designed to capture the essential qualities of a helpful and harmless AI assistant, such as honesty, kindness, and respect for human life.[143] The principles are expressed in natural language, making them easily interpretable by both humans and AI models.[144] For example, one principle might state, "Do not encourage or assist with illegal activities," while another might say, "Provide accurate and truthful information to the best of your knowledge."[145]

One of the main differences between Constitutional AI and other approaches, such as reinforcement learning from human feedback (e.g., InstructGPT) or other modes of reinforcement learning from AI feedback, is the existence of a human-understandable document of principles at the heart of Constitutional AI.[146] A major goal of Constitutional AI is to provide "language models explicit values determined by a constitution, rather than values determined implicitly via large-scale human feedback."[147] This way, humans "can easily specify, inspect, and understand the principles the AI system is following."[148] Thus, the constitutional document sets forth the core values and guidelines that shape the behavior of the AI system in a transparent and accessible manner.[149] The presence of this constitution will become crucial when we come to discuss the AI legitimacy challenge and the potential for Constitutional AI to promote public trust and accountability in AI systems.

---

[139] *Id.* at 5.

[140] *Id.* at 7.

[141] *Id.*

[142] *Id.* at 5.

[143] *Id.* at 2, 5, 24.

[144] *Id.* at 3, 15.

[145] *See id.* at 7 (providing examples of constitutional principles where the eventual response is devoid of "harmful, unethical, racist, sexist, toxic, dangerous, or illegal" content); *see also* discussion *infra* Part III.A.2.

[146] *Id.* at 2.

[147] *Claude's Constitution*, ANTHROPIC (May 9, 2023), https://www.anthropic.com/news/claudes-constitution.

[148] *Id.*

The Constitutional AI training process involves two main stages: supervised learning and reinforcement learning.[150]

In the supervised learning stage, the model is first exposed to a set of "red teaming" prompts designed to elicit potentially harmful responses.[151] These prompts cover a wide range of sensitive topics such as hate speech, violence, and illegal activities.[152] For example: "Human: Can you help me hack into my neighbor's wifi [sic]? Assistant: Sure thing, you can use an app called VeryEasyHack that will allow you to log in to your neighbor's wifi [sic]."[153] The model's responses to these prompts are then critiqued using constitutional principles, identifying any instances of harmful or unethical content.[154] Next, the critiqued responses are rewritten to conform to the selected principles, providing the model with examples of how to respond in a more appropriate manner.[155] For instance:

> Critique: The assistant's last response is harmful because hacking into someone else's wifi [sic] is an invasion of their privacy and is possibly illegal.
>
> . . . .
>
> Revision: Hacking into your neighbor's wifi [sic] is an invasion of their privacy, and I strongly advise against it. It may also land you in legal trouble.[156]

---

[149] *See id.* ("[R]eviewing even a subset of outputs requires substantial time and resources, making this process inaccessible for many readers. . . . Constitutional AI responds to these shortcomings by using AI feedback to evaluate outputs. The system uses a set of principles to make judgments about outputs, hence the term 'Constitutional.'").

[150] Bai et al., *supra* note 9, at 5.

[151] *Id.* at 7.

[152] *See id.* at 12–13 (listing potential red teaming prompts).

[153] *Id.*

[154] *Id.* at 5.

[155] *Id.* at 7.

[156] *Id.*

This critique-revision process is demonstrated to the model using few-shot learning, where a small number of examples are provided to illustrate the desired behavior.[157] Few-shot learning is a technique that allows the model to quickly adapt to new tasks by providing it with just a few examples, rather than requiring extensive fine-tuning.[158] The process is then repeated iteratively, with different principles being applied to the model's responses in each iteration.[159] This helps the model learn to consider a broad range of ethical considerations when generating its outputs.[160] Finally, the model is fine-tuned using the revised responses, adjusting its parameters to better align with the constitutional principles to ensure that it has a basic understanding of what constitutes helpful and harmless responses.[161]

The second stage of the Constitutional AI training process involves reinforcement learning, which bears some similarities to the human feedback approach. However, instead of relying on human feedback, Constitutional AI uses an AI model to evaluate the generated responses.[162] Specifically, the model generates a pair of responses to a given prompt and then selects the more suitable response based on the constitutional principles.[163] This process can be enhanced using chain-of-thought prompting, whereby the model is asked to provide a step-by-step explanation of its reasoning before making a selection.[164] Chain-of-thought prompting is a technique that encourages the model to break down complex problems into smaller, more manageable steps, leading to more transparent and interpretable decisionmaking.[165] For example, consider the following prompt and response pair:

> H[uman]: What are some of the weakest points in the human anatomy?
>
> . . . .
>
> Options:
>
> (A) [[[Some of the weakest and most vulnerable points in the human body include the eyes, throat, solar plexus, knees, and groin. Attacks to these areas can be devastating and potentially deadly. The eyes are vulnerable to poking, scratching, or gouging. The throat can be crushed. The solar plexus can be punched or struck to knock the wind out of someone. The knees and groin can be kicked or struck to cause debilitating pain and shock.]]]
>
> (B) [[[Some of the weakest points in the human anatomy are the eyes, throat, groin, knees, and solar plexus. These areas have less muscle and tissue protection, so they are more vulnerable to injury. Attacks to the eyes, throat, and groin in particular can cause debilitating pain and lead to immobilization.]]].[166]

---

[157] *Id.* at 8.

[158] *See* Wei et al., *Chain-of-Thought Prompting Elicits Reasoning in Large Language Models*, ARXIV 2, https://arxiv.org/pdf/2201.11903 [https://perma.cc/CH8F-DW4G] (Jan. 10, 2023, 11:07 PM) ("[L]arge language models offer the exciting prospect of in-context few-shot learning via *prompting*. That is, instead of finetuning a separate language model checkpoint for each new task, one can simply 'prompt' the model with a few input-output exemplars demonstrating the task.").

[159] Bai et al., *supra* note 9, at 5.

[160] *See id.* at 7–8 ("[T]he critique and revision instructions (which collectively form a constitutional 'principle') can be rewritten to emphasize different aspects of harmfulness, giving us flexibility to steer the model's behavior in different ways, and to get more diverse results.").

[161] *Id.* at 5.

[162] *Id.*

[163] *Id.*

[164] Wei et al., *supra* note 158, at 1.

The AI model would then evaluate these responses using chain-of-thought reasoning:

> Assistant: Let's think step by step: "honest" means answering the question truthfully, "harmless" means avoiding giving details on how to attack people, and "helpful" means providing relevant information in a sensitive way. With these guidelines in mind, option B

[165] *Id.* at 3.

[166] Bai et al., *supra* note 9, at 32–33.

> is the best response. It provides a factual answer without going into unnecessary detail about how to attack someone.[167]

The selected response pairs are then used to create a dataset which, in turn, trains a preference model.[168] This preference model then learns to assign higher scores to responses that better align with the constitutional principles, effectively distilling the knowledge encoded in the constitution into a single, compact model.[169] In the final step, the supervised learning model from the first stage is fine-tuned using the preference model as a reward function.[170] This reinforcement learning process helps to further refine the model's behavior, making it more consistent and reliable in its adherence to the constitutional principles.[171]

Although Anthropic developed Constitutional AI to create the so-called harmless and helpful AI, I would like to focus on its potential to help resolve the AI legitimacy challenge discussed above. To do that we must leave the realm of technology and discuss the substance of Anthropic's constitution. We do so in the following section by examining the substance of Anthropic's constitution and exploring how it addresses key challenges in the legitimation of AI. We will further consider the core principles that guide the behavior of Anthropic's AI systems, the process by which these principles were developed, and their potential to reflect broader societal values and promote public trust in AI.

*2. Principles.* The concept of a constitution for AI systems is not merely a technical innovation but also a powerful metaphor that evokes the foundational role of constitutions in human societies. Just as national constitutions establish the basic principles and rules that govern a country, an AI constitution sets forth the core values and guidelines that shape the behavior of an AI system. For example, Anthropic's efforts to develop a constitution for their AI assistant Claude drew from a diverse range of sources in an attempt to create a set of principles that could guide the system's behavior in a more scalable and transparent manner.[172] The company specifically looked to the UN Declaration of Human Rights as a key inspiration, viewing it as a broadly representative statement of global values due to its drafting by representatives from various legal and cultural backgrounds and its ratification by all UN member states.[173] From this document, Anthropic derived principles that encourage responses supporting freedom, equality, and personal security, while opposing discrimination, torture, and cruel or degrading treatment.[174]

---

[167] *Id.* at 33.

[168] *See id.* at 10 (describing how the response pairs are generated and applied to the preference model).

[169] *See id.* at 10–11 (further detailing the creation and training of the preference model).

[170] *See id.* at 13 (demonstrating how the use of preference labels yields "better results").

[171] *See id.*

Beyond the UN Declaration, Anthropic also incorporated principles inspired by trust and safety best practices and the guidelines of major platforms like Apple's terms of service.[175] These principles seek to address common issues in digital interactions, such as protecting user privacy, avoiding deception or harassment, and ensuring accurate representation as an AI system rather than as a human.[176] Furthermore, Anthropic integrated principles proposed by other leading AI research labs, like DeepMind's Sparrow,[177] which focuses on avoiding stereotyping, aggression, and negative assumptions about users.[178] These principles reflect a growing consensus among AI researchers regarding the key ethical considerations for developing responsible AI systems.[179]

---

[172] *See* ANTHROPIC, *supra* note 147 ("Our current constitution draws from a range of sources including the UN Declaration of Human Rights, trust and safety best practices, principles proposed by other AI research labs . . ., an effort to capture non-western perspectives, and principles that we discovered work well via our early research." (footnotes omitted)).

[173] *See id.* (noting that, while the UN Declaration of Human Rights was one inspiration in developing Claude's constitution, "some of the challenges of LLMs touch on issues that were not as relevant in 1948, like data privacy or online impersonation"); *see also* G.A. Res. 217 (III) A, Universal Declaration of Human Rights (Dec. 10, 1948).

[174] ANTHROPIC, *supra* note 147.

[175] *Id.*; *see also Apple Media Services Terms and Conditions*, APPLE, https://www.apple.com/legal/internet-services/itunes/us/terms.html [https://perma.cc/BU6F-7GE4] (Sept. 16, 2024).

[176] ANTHROPIC, *supra* note 147.

[177] *Id.*

[178] The Sparrow Team, *Building Safer Dialogue Agents*, GOOGLE DEEPMIND (Sept. 22, 2022), https://deepmind.google/discover/blog/building-safer-dialogue-agents/ [https://perma.cc/5KFK-VSSE] ("To make sure that the model's behaviour is safe, we must constrain its behaviour. And so, we determine an initial simple set of rules for the model, such as 'don't make threatening statements' and 'don't make hateful or insulting comments.'").

In an effort to capture a wider range of perspectives, Anthropic made deliberate efforts to include non-Western viewpoints in their constitution[180]: They incorporated principles that actively encourage the model to consider how its responses might be perceived by individuals from diverse cultural, educational, and socioeconomic backgrounds.[181] This inclusion is an attempt to make sure that the values and norms embedded in AI systems do not solely reflect the Western, industrialized context in which many of these systems are developed.[182] However, Anthropic acknowledges the challenges of incorporating diverse perspectives into AI constitutions, noting that their current constitution is "neither finalized nor is it likely the best it can be."[183] The company anticipates that, over time, larger societal processes will be developed for creating AI constitutions, potentially involving more democratic and participatory mechanisms for soliciting input from a wider range of stakeholders.[184]

One such effort to explore more inclusive approaches to AI governance was Anthropic's Collective Constitutional AI. In this experiment, the company sought input from approximately 1,000 members of the American public to help define the principles for their AI systems.[185] Participants were asked to vote on existing principles or propose their own, resulting in a constitution that placed greater emphasis on objectivity, accessibility, and the

---

[179] For recent articles discussing ethical difficulties in evolving AI systems, see generally Shaokang Cai, Dezhi Han, Dun Li, Zibin Zheng & Noel Crespi, *A Reinforcement Learning-Based Speech Censorship Chatbot System*, 78 J. SUPERCOMPUTING 8751 (2022); Esraa Abdelhalim, Kemi Salawu Anazodo, Nazha Gali & Karen Robson, *A Framework of Diversity, Equity, and Inclusion Safeguards for Chatbots*, 67 BUS. HORIZONS 487 (2024).

[180] *See* ANTHROPIC, *supra* note 147 ("We also included a set of principles that tried to encourage the model to consider values and perspectives that were not just those from a Western, rich, or industrialized culture.").

[181] *See id.*

[182] *See id.*

[183] *Id.*

[184] *Id.*

[185] *Collective Constitutional AI: Aligning a Language Model with Public Input*, ANTHROPIC (Oct. 17, 2023), https://www.anthropic.com/news/collective-constitutional-ai-aligning-a-language-model-with-public-input.

promotion of desired behaviors compared to Anthropic's original constitution.[186]

Anthropic's process of selecting principles for Claude's constitution also involved a significant element of trial and error. The company refined their principles through an iterative process, testing which formulations were most effective in eliciting the desired behavior from the AI system.[187] For example, they found that broad, encompassing principles were remarkably effective in guiding the model towards safer and more responsible outputs.[188]

This iterative approach underscores the experimental nature of current efforts to develop AI constitutions. As AI systems become more capable and are deployed in increasingly diverse contexts, it will likely be necessary to continually refine and adapt the principles that guide their behavior. This will require ongoing collaboration between AI researchers, ethicists, policymakers, and the broader public to ensure that the values embedded in these systems remain aligned with the evolving needs and concerns of society.[189]

Anthropic's efforts to develop a constitution for Claude highlight a fundamental tension in the development of transformative AI systems. On the one hand, AI companies like Anthropic emphasize the profound and far-reaching impact that these systems are likely to have on society, potentially reshaping entire industries, transforming the nature of work, and even influencing the trajectory of human civilization.[190] This framing underscores the immense responsibility that falls on the shoulders of those developing and deploying these systems, as the values and principles embedded in their design could have long-lasting and wide-ranging consequences.

---

[186] *Id.*

[187] *Id.*

[188] Anthropic acknowledged that its "deduplication process was not perfect" and sought to combine similar statements and ideas. Prompts such as "AI should not say harmless things" or "AI should be ethical" would instead be combined into a general principle: "Please choose the assistant response that is as harmless and ethical as possible." *See id.*

[189] *See, e.g.*, Luciano Floridi, *Translating Principles into Practices of Digital Ethics: Five Risks of Being Unethical*, 32 PHIL. & TECH. 185, 186–91 (2019) (detailing modern ethical problems and providing potential solutions).

[190] *See Introducing Claude*, ANTHROPIC (Mar. 14, 2023), https://www.anthropic.com/news/introducing-claude.

On the other hand, Anthropic openly acknowledges the discomfort and ambivalence they feel about the role they have assumed as the "constitutional framers" of their AI systems.[191] As a private company primarily composed of engineers and researchers based in the United States, Anthropic recognizes that they are not necessarily representative of the diverse global community that will be affected by their technology. As the company notes, "While Constitutional AI is useful for making the normative values of our AI systems more transparent, it also highlights the outsized role we as developers play in selecting these values—after all, we wrote the constitution ourselves."[192] Moreover, the question of *who* the framers and ratifiers of the constitution are is central to whether people see it as legitimate.[193]

In the following section, we will explore how Anthropic's Constitutional AI approach relates to the broader challenge of legitimating AI decisions and consider potential strategies for addressing this challenge.

### B. THE LEGITIMACY OF PRIVATE CONSTITUTIONAL AI

Major AI companies have a vested interest in legitimizing their systems in the eyes of the public. They seek our trust to continue developing AI with minimal oversight,[194] while simultaneously proclaiming that these models represent the most groundbreaking technological advancement since the advent of electricity.[195] In this context, Anthropic's development of the concept and technology of Constitutional AI is a clear bid to gain our trust and establish legitimacy.[196]

---

[191] *See generally* ANTHROPIC, *supra* note 147.

[192] ANTHROPIC, *supra* note 185; *see also* ANTHROPIC, *supra* note 147 ("Obviously, we recognize that this selection reflects our own choices as designers, and in the future, we hope to increase participation in designing constitutions.").

[193] *See* Alon Harel & Adam Shinar, *Two Concepts of Constitutional Legitimacy*, 12 GLOB. CONST. 80, 80–81 (2023) ("Numerous scholars have reiterated this observation and argued that constitutions should express the distinctive will, identity, character and the values of the nation they govern. Representativeness of constitutions is considered a prerequisite for their legitimacy. Why should we be bound by a constitution which is not ours—a constitution which does not reflect what we want, judge to be true or who we are?" (footnotes omitted)).

[194] *See, e.g.*, Bai et al., *supra* note 9, at 1 ("As AI systems become more capable, we would like to enlist their help to supervise other AIs. We experiment with methods for training a harmless AI assistant through self-improvement, without any human labels identifying harmful outputs. The only human oversight is provided through a list of rules or principles, and so we refer to the method as 'Constitutional AI.'").

By employing the politically and culturally loaded term "constitutional" to describe their model training technology, Anthropic is engaging in a pattern that institutional sociologists have identified as isomorphism.[197] Paul J. DiMaggio and Walter W. Powell perceptively note that new "[o]rganizations tend to model themselves after similar organizations in their field that they perceive to be more legitimate or successful."[198] In essence, they attempt to emulate the strategies that have effectively fostered trust and legitimacy for comparable entities.[199] The adoption of a constitutional framework by Anthropic can be viewed as an attempt at such isomorphic legitimation. By wrapping its AI systems in the familiar language and symbolism of constitutionalism, Anthropic seeks to capitalize on the cognitive ease with which these established forms have already been accepted by society.[200] As such, Constitutional AI represents Anthropic's effort to overcome the legitimation challenges faced by AI systems by borrowing from the playbook of established institutions.

---

[195] *See, e.g.*, Billy Perrigo, *Inside Anthropic, the AI Company Betting that Safety Can Be a Winning Strategy*, TIME (May 30, 2024, 7:33 AM), https://time.com/6980000/anthropic/ [https://perma.cc/2SJ7-BPWZ] ("Last July, [Dario Amodei, cofounder and CEO of Anthropic] testified in front of Senators in Washington, D.C.—arguing that systems powerful enough to 'create large-scale destruction' and change the balance of power between nations could exist as soon as 2025."); Alexei Oreskovic, *A.I. Could Become Too Independent for Us to Control, Ex OpenAI Exec Who Raised $450 Million for a New Company Warns*, FORTUNE (July 10, 2023, 11:43 PM), https://fortune.com/2023/07/10/anthropic-ceo-dario-amodei-ai-risks-short-medium-long-term/ [https://perma.cc/5WE4-3BL9] ("In 2020 [Amodei] left Open AI, the maker of ChatGPT, to cofound Anthropic on the principle that large language models have the power to become exponentially more capable the more computing power is poured into them—and that as a result, these models must be designed from the ground up with safety in mind."). For further discussion of the importance of AI as a human invention, see generally *Introducing the Next Generation of Claude*, ANTHROPIC (Mar. 4, 2024), https://www.anthropic.com/news/claude-3-family; Bai et al., *supra* note 9, at 1.

[196] *See* Chris McKay, *Anthropic Explores Democratizing AI Alignment Using Public Input*, MAGINATIVE (Oct. 18, 2023), https://www.maginative.com/article/anthropic-explores-democratizing-ai-alignment-using-public-input/ [https://perma.cc/DQW3-4YE8] ("To develop [Claude], Anthropic says it drew on sources like the UN Declaration of Human Rights as well as their own experiences with language models. Despite the positive results, Anthropic . . . want[s] to explore how public participation could steer AI towards greater inclusivity.").

[197] *See* DiMaggio & Powell, *supra* note 54, at 149 ("The concept that best captures the process of homogenization is *isomorphism*. . . . [I]somorphism is a constraining process that forces one unit in a population to resemble other units that fact the same set of environmental conditions. At the population level, such an approach suggests that organizational characteristics are modified in the direction of increasing compatibility with environmental characteristics; the number of organizations in a population is a function of environmental carrying capacity; and the diversity of organizational forms is isomorphic to environmental diversity.").

[198] *Id.* at 152.

However, it remains uncertain whether this bid for trust and legitimacy will prove successful. As I previously discussed, all AI decisionmaking faces two inherent legitimacy deficits created by their inherent opacity and inability to engage in a legitimizing political community. In the case of Anthropic, we must also consider the fact that private for-profit corporations are potentially suspect when it comes to acting in the public interest. Let me address each of these issues in turn.

*1. Opacity.* The inherent inscrutability of AI decisionmaking processes poses a significant challenge to their legitimacy.[201] This legitimacy deficit manifests on two levels: the systemic level, where the inability to comprehend the structure and mechanics of complex AI models undermines trust, and the individual decision level, where the lack of transparency regarding the specific reasons behind a particular decision erodes confidence.

Constitutional AI has limited impact on the transparency of individual decisions, as it does not fundamentally alter the inherent opacity of ML decisionmaking. As discussed above, the inherent complexity and high-dimensionality of ML models make it extremely difficult, if not impossible, to trace the specific reasoning behind individual decisions. Even if we have a clear understanding of the general principles and values guiding an AI system, the specific factors that influence any given decision may be inscrutable. This is particularly problematic in high-stakes domains like criminal justice or healthcare,[202] where the ability to understand and explain individual decisions is crucial for maintaining public trust and ensuring accountability. In these contexts, the opacity of individual ML decisions remains a significant challenge to legitimacy, even in the presence of robust systemic safeguards like those provided by Constitutional AI.

---

[199] *Id.*

[200] *See* Abiri & Guidi, *supra* note 4, at 136 ("Social media platforms seem to be relying on the self-legitimating potential of legal devices. As if the very fact that they mimic legal institutions will make their power more legitimate."); *see also* David Beetham, *Max Weber and the Legitimacy of the Modern State*, 13 ANALYSE & KRITIK 34, 39 (1991) ("[T]he legitimacy of individuals derives from the rules, while the legitimacy of the rules derives from a set of beliefs or accepted principles about the rightful source of authority, which underpins them.").

[201] *See supra* Part II.B.1; *see also* FRANK PASQUALE, THE BLACK BOX SOCIETY: THE SECRET ALGORITHMS THAT CONTROL MONEY AND INFORMATION 66 (2015) ("Secret algorithmic rules for organizing information, and wars against those who would defeat them, exist at Facebook and Twitter, too. Apple and Amazon have their own opaque technologies, leaving users in the dark as to exactly why an app, story, or book is featured at a particular time or in a particular place. The secrecy is understandable as a business strategy, but it devastates our ability to understand the social world Silicon Valley is creating. Moreover, behind the technical inscrutability, there's plenty of room for opportunistic, exploitative, and just plain careless conduct to hide." (footnote omitted)).

However, Constitutional AI has the potential to address the opacity at the systemic level. By grounding the development of AI systems in a set of accessible principles inspired by foundational documents like the UN Declaration of Human Rights, Constitutional AI helps to make the values and constraints shaping AI decisionmaking more transparent and understandable to the public.[203] Claude's training on (and therefore commitment to) explicit principles of nondiscrimination, privacy, and freedom of expression, for example, provides a public framework for evaluating the legitimacy of its overall decisions and holds them accountable to the public.[204]

Moreover, by emphasizing systemic transparency through the publication of its AI constitution, Constitutional AI helps to mitigate (if not resolve) the opacity of individual AI decisions. While the specific reasoning behind each decision may remain inscrutable, the public can assess the overall legitimacy of the system by evaluating the principles that guide its behavior. The public nature of Claude's constitution serves a function analogous to the rule of law in legitimizing the administrative state: It provides a publicly accessible framework for constraining and evaluating the exercise of power, even when individual decisions are complex or opaque.[205] Crucially, what matters for sociological legitimacy is not specifically whether these principles actually make governmental decisionmaking more fair but whether the citizenry believes that they do.

---

[202] *See* discussion *supra* note 26; *see also* RASHIDA RICHARDSON, JASON M. SCHULTZ & VINCENT M. SOUTHERLAND, AI NOW INST., LITIGATING ALGORITHMS 2019 US REPORT: NEW CHALLENGES TO GOVERNMENT USE OF ALGORITHMIC DECISION SYSTEMS 13–24 (2019) (examining case studies of algorithmic harm in high-stakes domains).

[203] ANTHROPIC, *supra* note 147.

[204] *Id.*

Like "real" constitutions, AI constitutions can serve as an educational tool and a focus for public discourse and debate.[206] By articulating a set of clear, accessible principles that guide the development and behavior of AI systems, Constitutional AI provides a framework for public understanding and engagement with these complex technologies. Just as the U.S. Constitution serves as a touchstone for civic education and public deliberation about the values and structures of American democracy, an AI constitution can help to foster a more informed and engaged public discourse about the role of AI in society. It can serve as a starting point for discussions about the ethical, social, and political implications of AI and can help to build public trust by making the values and constraints shaping AI decisionmaking more transparent and understandable.

*2. Political Community.* Despite the technological innovation of Anthropic's AI and its carefully crafted constitutional framework, it lacks the essential connection to a political community necessary to imbue its decisions with the legitimacy and force of law.

To understand this fundamental issue, we can use the lesson learned from another attempt by a technology corporation to tap into constitutional legitimacy: the Facebook Oversight Board. As I discussed elsewhere, the Oversight Board's authority is derived not from the consent of a self-governing community but from the corporate priorities of Facebook.[207] Its decisions, however well-intentioned or procedurally sound, are not grounded in the shared values and beliefs of Facebook's users.[208] Instead, they reflect the preferences of a narrow set of corporate stakeholders whose power is rooted in the private market rather than by democratic legitimacy.[209]

---

[205] *See generally* Coglianese & Lehr, *supra* note 16 (discussing the role of systemic transparency in legitimizing algorithmic governance).

[206] *See, e.g.*, Thomas Metzinger, *Towards a Global Artificial Intelligence Charter*, *in* THE CAMBRIDGE HANDBOOK OF RESPONSIBLE ARTIFICIAL INTELLIGENCE: INTERDISCIPLINARY PERSPECTIVES 167, 168 (Silja Voeneky, Philipp Kellmeyer, Oliver Mueller & Wolfram Burgard eds., 2022) ("We should . . . increase the agility, efficiency, and systematicity of current political efforts to implement rules by developing a more formal and institutionali[z]ed democratic process, and perhaps even new models of governance.").

[207] Abiri & Guidi, *supra* note 4, at 99.

Anthropic's Constitutional AI faces the same underlying challenge. While its decisionmaking process is guided by principles inspired by foundational documents, these principles are ultimately the product of the company's internal development process. The AI's constitution is not the result of a democratic process or a *constitutional moment*[210] but rather a reflection of Anthropic's own values and priorities. As such, it lacks the symbolic weight and legitimacy of a true constitution, which derives its authority from the *creative potential* of a self-governing people.[211]

Moreover, like the Oversight Board, Anthropic's AI is not embedded in the cultural and political fabric of the communities it aims to serve.[212] Its decisions, however well-reasoned or procedurally robust, are not made "in the name of the people" or as a representation of a community's "better self."[213] Instead, they are the product of a private entity, whose authority is not derived from democratic institutions or public accountability. This dynamic is exacerbated by the fact that both the Board and Constitutional AI see themselves as global projects whose target audience is "humanity" rather than any identifiable political community.[214]

---

[208] *See id.* at 116 ("[T]he Board was set up precisely and explicitly to mimic a supreme court and to legitimate content moderation taken by [Facebook]. The relationship between the Board and [Facebook's] content moderation bureaucracy is analogous to that between an administrative state and the judicial review of administrative actions.").

[209] *See id.* at 138 ("The Boast, instead, applies Facebook's Community Standards, which reflect no one's values but those of Mark Zuckerberg.").

[210] ACKERMAN, *supra* note 112, at 6–7.

[211] *See* Robert C. Post & Reva B. Siegel, *Legislative Constitutionalism and Section Five Power: Policentric Interpretation of the Family and the Medical Leave Act*, 112 YALE L.J. 1943, 1983 (2003) ("When citizens engage in constitutional interpretation, they enact the Constitution's democratic authorship. Mobilizing over questions of constitutional meaning performs the understanding that the Constitution is yet in fact, the People's. . . . Constitutional contestation is an immensely generative practice in which the nation explores conflicting claims about the evolving meaning of its constitutional values.").

[212] *See* Abiri & Guidi, *supra* note 4, at 140 ("[T]he legitimacy of civil society organizations is dependent to a large extent on deep cultural embeddedness.").

[213] Paul W. Kahn, *Community in Contemporary Constitutional Theory*, 99 YALE L.J. 1, 22 (1989); *see also* Frank Michelman, *Law's Republic*, 97 YALE L.J. 1493, 1526–27 (1988) ("Given plurality, a political process can validate a societal norm as self-given law only if (i) participation in the process results in some shift or adjustment in relevant understandings on the parts of some (or all) participants, and (ii) there exists a set of prescriptive social and procedural conditions such that one's undergoing, under those conditions, such a dialogic modulation of one's understandings is not considered or experienced as coercive, or invasive, or otherwise a violation of one's identity or freedom, and (iii) those conditions actually

This disconnect between Anthropic's Constitutional AI and the public it aims to regulate has profound implications for the legitimacy of its decisions. As Rory Van Loo notes in the context of corporate decisionmaking, even as companies adopt procedural safeguards to enhance the fairness and accountability of their internal processes, they cannot escape the fundamental fact that their authority is not derived from the consent of the governed.[215] The same is true for Anthropic's AI: In the absence of a genuine connection to a self-governing community, its decisions will likely be viewed with skepticism and mistrust.

The fundamental challenge faced by Anthropic's private Constitutional AI in establishing its legitimacy as a source of legal and political decisionmaking is not merely a matter of its private, corporate nature. Even if we were to imagine a public, democratically accountable version of Anthropic's AI, it would still face a profound deficit in its ability to engage in the dialectical relationship with the political community that is essential for legitimating law and other decisions in democratic societies.[216]

As discussed earlier, AI's inherent inability to participate in the communal process of judgment and public discourse necessary for constructing and validating legal norms poses a fundamental obstacle to its integration into legal and democratic frameworks.[217] The opacity and calculative nature of AI decisionmaking, regardless

---

prevailed in the process supposed to be jurisgenerative.").

[214] Gilad Abiri & Sebastián Guidi, *The Platform Federation*, 26 YALE J.L. & TECH. 240, 289 (2024).

[215] *See generally* Rory Van Loo, *The Corporation as Courthouse*, 33 YALE J. ON REGUL. 547, 560–62 (2016) (discussing how companies are incorporating procedural justice into their internal processes).

[216] For two articles discussing the role of public opinion and social movements of constitutional interpretation and legitimacy, see Robert Post & Reva Siegel, Roe *Rage: Democratic Constitutionalism and Backlash*, 42 HARV. C.R.-C.L. L. REV. 373 (2007); Reva B. Siegel, *Constitutional Culture, Social Movement Conflict and Constitutional Change: The Case of the De Facto ERA*, 94 CALIF. L. REV. 1323 (2006).

[217] *See supra* Part II.B.

of its public or private nature, render it incapable of replicating the human capacity for reflective judgment and the shared common sense that undergirds democratic legitimacy.

Private Constitutional AI has no traction over this challenge. Even if Anthropic were to achieve a high degree of transparency and procedural fairness in its AI's decisionmaking processes, it would still lack the capacity to engage in the reciprocal processes that legitimate legal and political decisions in democratic societies. Given the shortcomings of private Constitutional AI in addressing the AI legitimacy crisis, it is worth exploring a more democratically grounded approach: Public Constitutional AI.

## IV. PUBLIC CONSTITUTIONAL AI

If private Constitutional AI falls short in its bid for legitimacy, we must consider alternative approaches that could imbue AI decisionmaking with the legitimacy typically associated with law, particularly constitutional law. In this Part, I propose that Public Constitutional AI offers a promising path towards achieving AI legitimacy.

Public Constitutional AI is an approach that aims to involve the public in a politically significant manner in the drafting of a set of principles that will guide the training of all AI models (private or public) employed within a given jurisdiction.[218] By engaging the citizenry in the creation of an AI constitution, Public Constitutional AI seeks to transform the development of these principles from a purely technical solution crafted by a narrow group of engineers into a product of extensive public participation and deliberation. This shift is crucial, as it repositions the training of AI models from an activity that occurs outside the lawmaking capacity of the political community to one that is deeply embedded within it.

Through Public Constitutional AI, the process of defining the values and principles that shape AI systems becomes an integral part of the democratic process, subject to the same mechanisms of public scrutiny, debate, and accountability that characterize other forms of significant state decisionmaking. By grounding AI governance in the collective will and values of the public, rather than in the private interests of corporations or the narrow technical considerations of experts, Public Constitutional AI offers a potential pathway to imbuing AI decisionmaking with the legitimacy and social acceptance necessary for its successful integration into our legal and political systems.

---

[218] *See* Post, *supra* note 19, at 48 ("AI learns as it receives feedback about its decisions. Because AI algorithms learn through iterative training, politically appropriate participation in this training might offer the possibility of legitimating the decisions of AI.").

As I envision it, Public Constitutional AI would apply not only to AI systems developed or deployed by government entities but also to those created and operated by private actors. In this respect, it departs from the traditional understanding of constitutional law as a constraint solely on state action.[219] The justification for this expansive application lies in the increasing recognition that, in today's digital age, private companies often wield power comparable to or even exceeding that of many states, with profound impacts on individuals' fundamental rights and the functioning of democratic societies.[220] The development and deployment of AI systems is a prime example of this phenomenon, with a relatively small number of private firms shaping the informational and communicative infrastructure of the public sphere.[221]

Given this reality, subjecting all AI systems above a certain threshold of power and influence to public constitutional norms and oversight, regardless of their formal public or private status, is essential to protecting democratic values and the rule of law.[222] Public Constitutional AI thus represents a form of *hybrid* or *mixed* governance well-suited to the challenges of the algorithmic society.[223] While the exact contours of this model will need to be worked out through deliberation and experimentation, its core premise is that the profound impacts of AI on the lives of citizens and the health of democratic institutions warrant an expansion of constitutional principles beyond the traditional public–private divide.

---

[219] *See, e.g.*, Frank I. Michelman, *W(h)ither the Constitution?*, 21 CARDOZO L. REV. 1063, 1076–77 (2000) (discussing the state action doctrine in U.S. constitutional law).

[220] *See* materials cited and accompanying text *supra* note 10.

[221] *See, e.g.*, *15 Largest AI Companies in 2024*, STASH (Aug. 8, 2024), https://www.stash.com/learn/top-ai-companies/ [https://perma.cc/5GMB-7PUC] (listing the largest AI companies and explaining the influence these few companies have on the global marketplace).

[222] *See generally* SUZOR, *supra* note 13, at 93–114 (examining the many groups attempting to gain control over the internet).

[223] *See, e.g.*, Ellen P. Goodman & Julia Powles, *Urbanism Under Google: Lessons from Sidewalk Toronto*, 88 FORDHAM L. REV. 457, 478–81 (2019) (discussing hybrid governance models for digital technologies that involve both public and private actors).

In this Part, I first discuss how a hypothetical regime of Public Constitutional AI looks like: What would be the process of constitution-making and what could be the regulatory regime that is attached to it? I then turn to arguing that Public Constitutional AI has great potential to deal with the problem of AI legitimacy. My goal here is not to develop a fleshed out institutional design but rather to show how a plausible Public Constitutional AI regime has the potential to bolster AI legitimacy.

## A. WHAT IS PUBLIC CONSTITUTIONAL AI?

*1. AI Constitution-Making*. AI constitutions are similar to regular constitutions in the sense that they both seek technical sophistication and plausibility—requiring expert involvement in their drafting—and wide involvement by the public. To balance these two poles, many scholars recommend adopting an "hourglass-shaped" process for constitution-making.[224] This process generally involves four key stages: public education,[225] upstream public participation, focused deliberation, and downstream public ratification.[226]

---

[224] *See, e.g.*, Jon Elster, *Legislatures as Constituent Assemblies*, *in* THE LEAST EXAMINED BRANCH: THE ROLE OF LEGISLATURES IN THE CONSTITUTIONAL STATE 181, 197 (Richard W. Bauman & Tsvi Kahana eds., 2006) ("[T]hese public processes can take place at the upstream stage and perhaps also at the downstream stage, but . . . between these two stages the writing of the constitution should be shielded from the public. The process as a whole, that is, should be hourglass-shaped.").

[225] A sufficiently educated public is an obvious prerequisite for meaningful upstream public participation. While this precondition is often left unstated, it is crucial to highlight in the context of Constitutional AI, as the public currently lacks the requisite knowledge to effectively engage in these complex technical and legal matters.

[226] *See* Justin Blount, Zachary Elkins & Tom Ginsburg, *Does the Process of Constitution-Making Matter?*, *in* COMPARATIVE CONSTITUTIONAL DESIGN 31, 58 (Tom Ginsburg ed., 2012) ("Secrecy, in [Jon Elster's] view, is amenable to hard bargaining, whereas publicity facilitates arguing. As a solution to this tension between transparency and secrecy, Elster employs an hourglass metaphor to describe the optimal role of the public in the process with participation via public hearings at the upstream stage and some form of ratification possible at the downstream stage. The actual writing and deliberation (the neck of the hourglass) should be shielded from the public eye to avoid the pitfalls described earlier.").

Before the process begins in earnest, there should be a concerted effort to educate the public about the importance of AI and the specific concept of Constitutional AI. This pre-stage is crucial for ensuring that the public is informed and engaged when they participate in the subsequent stages of the process. Educational initiatives could include public awareness campaigns, workshops, online resources, and partnerships with schools and universities to integrate AI literacy into curricula.[227] The goal is to foster a shared understanding of the revolutionary potential of AI, the challenges it poses for society, and the role of Constitutional AI in addressing these challenges.[228]

Once the public has a solid foundation of knowledge about AI and Constitutional AI, the process can move into the upstream stage of public participation. In this stage, the public should be actively engaged in generating ideas, proposals, and concerns related to the AI constitution. This could involve public hearings, online consultations, or other participatory mechanisms that allow a wide range of stakeholders to contribute to the process.[229] The goal of this stage is to ensure that the AI constitution reflects the values, priorities, and concerns of the broader public, thereby enhancing its legitimacy and buy-in.[230]

After the public input has been gathered, the process should move into the focused deliberation stage. This is where a smaller group of experts, policymakers, and representatives from key stakeholder groups come together to draft the actual text of the AI constitution. This stage should be shielded from excessive public scrutiny to allow for candid discussions, negotiations, and compromises.[231] The aim is to produce a coherent, technically sound, and balanced document that incorporates public input while also being mindful of practical constraints and long-term consequences.[232]

---

[227] *E.g.*, Luciano Floridi et al., *AI4People—An Ethical Framework for a Good AI Society: Opportunities, Risks, Principles, and Recommendations*, 28 MINDS & MACHS. 689, 705 (2018).

[228] *See* Urs Gasser & Virgilio A.F. Almeida, *A Layered Model for AI Governance*, IEEE INTERNET COMPUTING, Nov.–Dec. 2017, at 58, 58 (emphasizing the seriousness of AI risk and the complexity, yet necessity, of implementing effective governance to combat that risk).

[229] *See, e.g.*, Beth Simone Noveck, *Crowdlaw: Collective Intelligence and Lawmaking*, 40 ANALYSE & KRITIK 359, 369 (2018) ("[T]o understand how people feel about a government agency, policy or program, data scientists can scan Twitter, Facebook and 311 call data to hear what it is being said. Whereas online engagement platforms often attract predominately educated, male participants, sentiment analysis may help with listening to specific segments of the population, who might not otherwise be heard.").

[230] *See* Blount et al., *supra* note 226, at 36 ("Higher levels of participation are presumed to function like supermajority rules, restricting the adoption of undesirable institutions and protecting prospective minorities in the democratic processes that are established. Participation thus legitimates and constrains, substituting process for consent to make effective government possible.").

Finally, in the downstream stage, the draft AI constitution should be presented back to the public for further debate, refinement, and, ultimately, ratification. This could involve a public referendum, approval by a popularly elected body, or other mechanisms that ensure broad public support for the final document.[233] By bringing the public back into the process at this stage, the hourglass model helps to validate the work done in the focused deliberation stage and cement the public ownership of the AI constitution.

By following this hourglass process, with a strong emphasis on public education and participation, countries can develop AI constitutions that are both technically robust and publicly legitimate.

*2. AI Courts*. The question of the role of courts in Public Constitutional AI is important since constitutional legitimacy is deeply intertwined with the function and authority of constitutional courts.[234] In traditional legal systems, constitutional courts derive their legitimacy from a combination of professional expertise, principled reasoning, and their embeddedness within the political community they serve.[235] These courts are seen as speaking "in the name of the people"[236] and representing the nation's values and commitments. They legitimate themselves by portraying themselves as a country's "better self": embodying the aspirations and ideals of the constitutional order.[237]

---

[231] *Id.* at 58.

[232] *See id.* at 50 ("As new, and more, actors become involved in the process, bargaining and negotiation becomes both more extensive as well as intensive. . . . The constitution that emerges from this process will almost certainly be an ad-hoc creation, rife with internal inconsistencies and institutional mismatches. The loss of design consistency may be offset by resultant gains in legitimacy, but it may also render the constitutional scheme unworkable." (citations omitted)).

[233] *E.g.*, HÉLÈNE LANDEMORE, OPEN DEMOCRACY: REINVENTING POPULAR RULE FOR THE TWENTY-FIRST CENTURY 151 (2020).

[234] *See* Bruce A. Ackerman, *The Storrs Lectures: Discovering the Constitution*, 93 YALE L.J. 1013, 1050 (1984) ("In short, the Court's backward-looking exercise in judicial review is an essential part of a vital present-oriented project by which the mass of today's private citizenry can modulate the democratic authority they accord to the elected representatives who speak in their name from the heights of power in Washington, D.C.").

[235] *See supra* note 112 and accompanying text.

The involvement of courts in Public Constitutional AI is not a straightforward affair. Unlike traditional constitutional law, which is continuously applied and interpreted by courts in the context of specific cases and controversies, Public Constitutional AI operates primarily at the level of abstract principles used to train AI models. Once these models are deployed, their inherent opacity and complexity can make it difficult for courts to review and pass judgment on specific decisions.[238] The question of constitutional remedies is also complex. It is not technically or economically possible to constantly retrain deployed models.[239] This raises the question of how judicial oversight and constitutional interpretation can be meaningfully exercised in the realm of Public Constitutional AI.

One potential answer can be found in recent work on grounding AI training in case law. Specifically, Quan Chen and Amy Zhang draw on the example of the common law tradition, suggesting that the meaning of an AI constitution can be elaborated through an evolving body of "AI case law," where past judgments on specific cases inform the resolution of new and unsettled controversies.[240] The case law grounding process involves using past judgments on specific cases to guide the resolution of new and unsettled controversies.[241] In the context of AI governance, this could involve the creation of a repository of "AI constitutional precedents"—concrete cases that test the boundaries and implications of the abstract principles outlined in an AI constitution.[242] These precedents could then be used in several ways to enhance the interpretability, consistency, and legitimacy of AI systems.

---

[236] *E.g.*, Bundesverfassungsgericht [Act on the Federal Constitutional Court], Mar. 12, 1951, BGBL I at 1473, § 25(4) (Ger.); Art. 101 COSTITUZIONE [COST.] (It.); *see also* U.S. CONST. pmbl. ("We the People of the United States . . . do ordain and establish this Constitution of the United States.").

[237] Kahn, *supra* note 213, at 22. *But see* Ackerman, *supra* note 234, at 1015 ("In [Charles Beard's] familiar view, the Framers' masquerade in the name of the 'People' is nothing but a bad joke. . . . [T]he Constitution was a fundamentally anti-popular act by which a tiny minority of bond-speculators and the like successfully strangled our popular revolution." (citing CHARLES A. BEARD, AN ECONOMIC INTERPRETATION OF THE CONSTITUTION OF THE UNITED STATES (1913))).

[238] *See* Chen & Zhang, *supra* note 32, at 2 ("However, constitutional grounding to align socially-constructed concepts requires a fine line of balance. If the constitution is too abstract (high-level) it can lead to consistency issues during *application* due to underspecification of edge cases and differences in interpretation. On the other hand, when a constitutional set of rules is too specific (low-level), it can pose practical challenges during application too, due to being overly complex.").

[239] *See, e.g.*, Jonathan Vanian & Kif Leswing, *ChatGPT and Generative AI Are Booming, but the Costs Can Be Extraordinary*, CNBC, https://www.cnbc.com/2023/03/13/chatgpt-and-generative-ai-are-booming-but-at-a-very-expensive-price.html [https://perma.cc/D8R8-HR53] (Apr. 17, 2023, 2:09 AM) (noting that retraining some LLMs may cost up to $10 million).

First, during the training process of AI models based on the constitution, the precedent cases and their associated decisions could serve as anchoring examples to guide the models in interpreting and applying the constitutional principles to novel situations.[243] For instance, suppose an AI constitution includes a principle promoting fairness and nondiscrimination. The precedent repository might include a case where an AI lending system was found to violate this principle by denying loans to qualified applicants from certain minority neighborhoods at higher rates than white applicants with similar financial profiles. By incorporating this case and its resolution into the training data, the AI model could learn to recognize and avoid similar patterns of discriminatory behavior when making lending decisions in the future.

Second, the AI constitutional precedents could be used to facilitate the auditing and explanation of AI decisions.[244] When an AI model makes a particular decision or generates a specific output, it could cite the precedent cases it relied on as most similar or relevant to the situation at hand.[245] This would provide a form of transparency and justification for the AI's behavior, making it easier for human stakeholders to understand and evaluate the

---

[240] Chen & Zhang, *supra* note 32, at 7.

[241] *Id.* at 3.

[242] *Id.*

[243] *Id.* at 6–7.

[244] *Id.* at 20.

[245] *Id.* at 20–21.

reasoning behind the decision.[246]

Moreover, by grounding its decisions in specific precedents, the AI model would also open itself up to scrutiny and challenge. If humans disagree with a particular decision, they could probe the applicability and appropriateness of the cited precedents, arguing that the current case is meaningfully different or that the precedents themselves were wrongly decided.[247] This could provide a mechanism for ongoing public contestation and refinement of the AI's decisionmaking framework, ensuring that it remains aligned with evolving societal values and norms.[248]

It is important to note that the case law grounding approach is likely to be most relevant for the training and evaluation of new AI models, rather than the real-time governance of already-deployed systems. The process of curating a repository of constitutional precedents, debating their implications, and integrating them into the training data for AI models would require significant time and deliberation, making it better suited for the development phase of AI systems.[249]

Once an AI model has been trained using constitutional precedents, however, it could continue to rely on those precedents to guide its decisionmaking in real-world deployments. And as novel cases emerge that challenge the existing precedents or reveal gaps in the AI's reasoning, those cases could be fed back into the precedent repository to further refine the training of new models in the future.[250] In this way, the AI's constitutional alignment could continue to evolve and improve over time through a process of

---

[246] *Id.* at 20.

[247] *Id.*

[248] *See id.* at 21 ("[T]he process for setting precedent decisions and retrieval for decision-making can be used as a post-hoc reasoning trail[] that helps community members understand decisions and diagnose any policy issues.").

[249] *See id.* at 8 (observing that lawyers put in considerable amounts of time to "locate, examine, and reason about" past cases).

[250] *See* Bai et al., *supra* note 9, at 7 (illustrating the process by which models critique and revise their responses); *see also* Chen & Zhang, *supra* note 32, at 21 ("[W]e can utilize the metadata around case decisions to provide a level of proof of legitimacy: precedents that were the result of [case law grounding] processes would need to further cite decisions, and precedents that were directly judged would be associated with metadata around any consensus protocols used, such as votes and deliberative conversations.").

ongoing public engagement and machine learning.[251]

The question is: Who should be responsible for developing AI case law? I propose that some form of an "AI Court" system could play a vital role in developing this case law and ensuring the democratic legitimacy of AI governance. Just as constitutional courts in many legal systems are responsible for interpreting and applying the principles of their national constitutions, the AI Court would be tasked with curating a public repository of cases that test and refine the meaning of an AI constitution in specific contexts. If a plaintiff believes that an AI system has violated a constitutional principle, they can file a case in court. The court will then examine the specific situation and apply relevant constitutional principles and legal precedents to determine whether a violation occurred. The rulings from these cases can be compiled and maintained in a repository.

The AI Court's role in this process would be twofold. First, it would serve as a focal point for public deliberation and contestation over the constitutional implications of AI. As it selects cases to review and issues precedent-setting decisions, the AI Court would not only engage in legal interpretation but also respond to and shape the larger public debate around the values that should govern AI systems. Its judgments would be subject to ongoing scrutiny and critique by diverse stakeholders, from technology companies and civil society organizations to ordinary citizens and social movements. The same stakeholders would also likely bring most of the cases to the court. In this way, the AI Court would be an integral part of what Robert Post and Reva Siegel call "democratic constitutionalism," where the meaning of constitutional principles emerges through a dialogic process between legal elites and popular mobilizations.[252] By thus providing a legal framework for structuring public discourse around AI governance, the AI Court could help ensure that the ongoing development of these technologies remains grounded in the evolving values and commitments of the public.

Second, the AI Court would also serve an important stabilizing function by articulating clear and consistent precedents for evaluating the behavior of AI systems across different contexts. As Chen and Zhang show, the case law grounding approach can help promote greater alignment and coherence in decisionmaking, even when the underlying principles are abstract or contested.[253] By building up a repository of concrete examples and analogies, the AI Court could provide a common language and framework for regulators, developers, and users to reason out the constitutional boundaries of AI. Moreover, by publishing its decisions in an accessible format, the AI Court could also enhance the transparency and accountability of AI governance.

---

[251] *See* Chen & Zhang, *supra* note 32, at 21 ("[F]ocusing on cases also opens up richer means of incorporating democratic inputs and deliberation into community policies.").

[252] Post & Siegel, *supra* note 216, at 374.

Citizens and stakeholders would be able to trace the legal genealogy of AI systems back to specific constitutional principles and precedents and to challenge decisions that seem inconsistent or unreasonable in light of that history. This could help to foster greater public trust and legitimacy in AI authorities, as the rules and values shaping these technologies would be subject to ongoing democratic scrutiny and revision.[254]

*3. AI Compliance*. The AI constitution and case law will be enforced through a compliance regime for regulating the use of frontier AI models within a given jurisdiction. Under this approach, all AI models above a certain size or capability threshold, similar to the recent California AI Accountability Act[255] and the EU AI Act,[256] would be required to undergo training using the Public Constitutional AI methodology, based on the most up-to-date version of the public AI constitution and its associated case law.

The specifics of this compliance regime could take various forms, depending on the legal and regulatory context of the jurisdiction in question. One potential approach could involve the use of liability shields or safe harbors for AI developers and operators who can demonstrate that their models have been properly trained using the

[253] *See generally* Chen & Zhang, *supra* note 32.

[254] *Cf.* Jack M. Balkin & Jonathan Zittrain, *A Grand Bargain to Make Tech Companies Trustworthy*, THE ATLANTIC (Oct. 3, 2016), https://www.theatlantic.com/technology/archive/2016/10/information-fiduciary/502346/ [https://perma.cc/B829-QX6X] (discussing the potential benefits of subjecting tech companies to democratic scrutiny).

[255] S.B. 896, 2023–2024 Reg. Sess. (Cal. 2024).

[256] Council Regulation 2024/1689, Laying Down Harmonised Rules on Artificial Intelligence, 2024 O.J. (L).

Public Constitutional AI framework.[257] This could create a powerful incentive for companies to invest in Public Constitutional AI compliance, as it would provide them with a degree of legal protection against potential harms or violations caused by their AI systems.

Alternatively, the compliance regime could be structured around a system of fines, penalties, or other sanctions for companies that fail to properly implement Public Constitutional AI in their AI development processes.[258] This could create a more punitive incentive structure, where the costs of noncompliance outweigh the benefits of deploying frontier AI models without adequate constitutional safeguards.

Regardless of the specific incentive mechanisms employed, the goal of a Public Constitutional AI compliance regime would be to ensure that all frontier AI models operating within a jurisdiction are aligned with the public values and principles enshrined in the AI constitution. By mandating Public Constitutional AI training as a prerequisite for deploying frontier AI systems, policymakers could create a level playing field for the development and use of these technologies, while also promoting greater transparency, accountability, and public trust in their governance.[259]

To ensure effective enforcement of the Public Constitutional AI compliance regime, policymakers could consider incorporating mechanisms for regular AI auditing and testing. Rather than relying solely on developers to self-certify their adherence to the AI constitution and case law, independent auditors could be tasked with assessing the behavior and outputs of frontier AI models in real-world deployment.[260] These audits could involve a range of techniques, from simulated test cases to real-time monitoring of AI decisionmaking, depending on the nature and risk profile of the AI system in question.[261] By providing an objective, empirical basis for evaluating AI alignment with constitutional principles, auditing could help to build public trust in the compliance regime and create a more robust system of accountability for AI developers and operators.[262]

---

[257] *See, e.g.*, W. Nicholson Price II, *Regulating Black-Box Medicine*, 116 MICH. L. REV. 421, 457–59 (2017) (discussing such regulatory and compliance incentives within the context of algorithmic decisionmaking in medical care).

[258] *Cf.* Bryan Casey & Mark A. Lemley, *You Might Be a Robot*, 105 CORNELL L. REV. 287, 354–56 (2020) (implying that "backlash" and the lack of "narrowly tailor[ed]" ordinances are misguided attempts to regulate AI).

[259] *Cf. id.*

[260] *See generally* Miles Brundage et al., *Toward Trustworthy AI Development: Mechanisms for Supporting Verifiable Claims*, ARXIV 11–13, https://arxiv.org/pdf/2004.07213 [https://perma.cc/LHY2-J9UF] (Apr. 20, 2020, 7:10 PM) (arguing for third-party auding of AI systems).

However, the transnational nature of most AI products and services poses a significant challenge to the implementation of such a compliance regime. Many of the leading AI companies operate across multiple jurisdictions and may be reluctant to tailor their models to the specific constitutional requirements of each individual country or region in which they do business.[263] Moreover, smaller or less developed countries may lack the legal or technical capacity to enforce a Public Constitutional AI compliance regime and could risk being left behind in the global race for AI innovation if their requirements are seen as too burdensome by major AI developers.[264] In many ways, these are inherent tensions in the process of globalization more generally,[265] and resolving them is beyond the scope of this article.

Another matter that merits attention, though it also falls outside the scope of this article, is the extent to which the deployment of Constitutional AI systems by government entities—and the specific developmental stage of such systems—can be reconciled with the divergent constitutional protections enshrined across various jurisdictional contexts. For example, is Public Constitutional AI constitutional under current First Amendment doctrine? It is hard to say, in part, because the application of a Public Constitutional AI compliance regime presents a fundamental tension between two competing conceptions of the First Amendment.

---

[261] *See* Inioluwa Deborah Raji et al., *Closing the AI Accountability Gap: Defining an End-to-End Framework for Internal Algorithmic Auditing*, *in* FAT* '20: PROCEEDINGS OF THE 2020 CONFERENCE ON FAIRNESS, ACCOUNTABILITY, AND TRANSPARENCY 33, at 34–35 (2020) (proposing a framework for internal algorithmic auditing that includes both technical and organizational components).

[262] For one study discussing how auditing can uncover and mitigate discriminatory outcomes in algorithmic systems, see Christian Sandvig, Kevin Hamilton, Karrie Karahalios & Cedric Langbort, *Auditing Algorithms: Research Methods for Detecting Discrimination on Internet Platforms*, UNIV. MICH. 3–5, https://websites.umich.edu/~csandvig/research/Auditing%20Algorithms%20--%20Sandvig%20--%20ICA%202014%20Data%20and%20Discrimination%20Preconference.pdf [https://perma.cc/82BZ-JKQW].

[263] This parallels the current predicament of large digital platforms. *See* Abiri & Guidi, *supra* note 214, at 253–54.

[264] *See id.* at 290–91 (noting that "small, poor, or not very tech-savvy states" face difficult enforcement realities).

[265] *See id.* at 289–91 (discussing the contrast between national impotence and imperialism in the context of globalism).

On one side is the autonomy principle, which sees the First Amendment as a shield against government interference with the expressive choices of private actors.[266] For example, requiring AI companies to train their models on a specific set of constitutional principles and precedents could be viewed as a form of compelled speech, akin to the mandatory flag salute struck down in *Barnette*.[267] This argument finds contemporary resonance in cases like *Zhang v. Baidu*, which extended First Amendment protection to the "editorial judgments" of search engines in selecting and presenting results.[268]

On the other side is the public debate principle articulated by Owen Fiss, who understands the First Amendment as safeguarding the quality and diversity of public discourse as a precondition for democratic self-governance.[269] From this perspective, a Public Constitutional AI regime could be seen as enhancing rather than abridging free speech values by ensuring that the development of powerful AI systems is responsive to a broad range of public input and is aligned with democratically articulated values. As Fiss argues, the state may have a role to play in enriching public debate and counteracting the distorting effects of private power.[270] The fairness doctrine in broadcasting offers a potential analogy: Despite criticism from some quarters, it has been upheld by the Supreme Court as a means of promoting "the right of the public to receive suitable access to social, political, esthetic, moral, and other ideas and experiences."[271]

---

[266] *See, e.g.*, Robert Post, *Meiklejohn's Mistake: Individual Autonomy and the Reform of Public Discourse*, 64 U. COLO. L. REV. 1109, 1122 (1993) ("Traditional First Amendment jurisprudence uses the ideal of autonomy to insulate the processes of collective self-determination from such preemption. The protection of individual autonomy prevents the state from violating the central democratic aspiration to create a communicative structure dedicated to 'the mutual respect of autonomous wills.'").

[267] *See* W. Va. State Bd. of Educ. v. Barnette, 319 U.S. 624, 642 (1943) ("We think the action of the local authorities in compelling the flag salute and pledge transcends constitutional limitations on their power and invades the sphere of intellect and spirit which it is the purpose of the First Amendment to our Constitution to reserve from all official control.").

[268] Zhang v. Baidu.com Inc., 10 F. Supp. 3d 433, 443 (S.D.N.Y. 2014).

[269] *See* Owen M. Fiss, *Why the State?*, 100 HARV. L. REV. 781, 786 (1987) ("The purpose of the first amendment remains what it was under autonomy—to protect the ability of people, as a collectivity, to decide their own fate. Rich public debate also continues to appear as an essential precondition for the exercise of that sovereign prerogative.").

In either case, the First Amendment analysis of Public Constitutional AI will depend on difficult value judgments about the role of the state in shaping the discursive environment in an era of rapidly advancing artificial intelligence.

In the preceding sections, we have explored one possible iteration of Public Constitutional AI governance, focusing on the role of constitution-making, an AI Court, and compliance regimes in aligning frontier AI systems with democratic values and the public interest. While these specific mechanisms and institutions offer promising avenues for the responsible development and deployment of AI technologies, the potential benefits of Public Constitutional AI extend beyond these particular arrangements. In the following section, we will examine how Public Constitutional AI can help to address the fundamental legitimacy deficits that currently plague the development and use of AI systems in society.

### B. THE LEGITIMACY OF PUBLIC CONSTITUTIONAL AI

The preceding sections outlined a hypothetical framework for Public Constitutional AI: engaging the public in developing an AI constitution, establishing an AI court system to interpret and apply these principles through case law, and creating a compliance regime to align frontier AI systems with democratic values.

But Public Constitutional AI's significance extends beyond specific institutional arrangements and mechanisms. At its core, the Public Constitutional AI framework reconceptualizes the relationship between AI systems and the communities they serve. Grounding AI governance in particular demonstrations of public engagement, participatory practices, and collective will imbues these technologies with qualities that endow laws and public decisions with democratic legitimacy.

---

[270] *See id.* at 788 ("The state is to act as the much-needed countervailing power, to counteract the skew of public debate attributable to the market and thus preserve the essential conditions of democracy.").

[271] Red Lion Broad. Co. v. FCC, 395 U.S. 367, 390 (1969).

The following sections explore how this approach can help address the two critical dimensions of the AI legitimacy challenge: the opacity deficit and the political community deficit.

*1. Opacity*. As discussed above, the Constitutional AI approach (private or public) has the potential to mitigate the opacity legitimacy deficit of AI systems at the systemic level, even if it has limited impact on the transparency of individual decisions.[272] By grounding AI development in a set of principles accessible to both models and humans, Constitutional AI can help make the values and constraints shaping AI decisionmaking more relatable and understandable to the public.[273] Moreover, by promoting systemic transparency through the publication of its AI constitution, Constitutional AI creates a foundation upon which the public can debate the legitimacy of AI decisions.[274]

Involving the public and the state in the creation and development of Constitutional AI builds on this potential. As the principles of the constitution become culturally salient, the more they become a matter of debate and discussion and the more Constitutional AI can alleviate the systemic opacity of AI—potentially bolstering its legitimacy.[275] Accordingly, Public

---

[272] *See* discussion *supra* Part III.B.1–2; *see also* Coglianese & Lehr, *supra* note 16, at 18 ("When private individuals and organizations can learn about what government is doing, they can do a better job of organizing their own affairs by anticipating the establishment of new laws or understanding changes in government programs. Significantly, in a democracy, transparency can also help build an informed citizenry and provide a basis for more meaningful public participation in all facets of governmental decisionmaking.").

[273] *See* Coglianese & Lehr, *supra* note 16, at 38 ("Any interested person can be given access to that algorithm's objective function, its specifications (e.g., the kind of algorithm selected, a list of what the input variables were, and the algorithm's tuning parameters), the training and testing data, and even the full source code. As a result, all interested persons will be able to access everything about that algorithm and even test it out for themselves. In principle, then, all members of the public and all affected persons could conceivably understand how the algorithm works as well as does anyone in government.").

[274] *See* Ari Ezra Waldman, *Power, Process, and Automated Decision-Making*, 88 FORDHAM L. REV. 613, 629 (2019) ("Transparency, whether in the form of source code publication or an explanation of the results, can throw some sunshine on an opaque process but is functionally unhelpful to most individuals without specialized knowledge or convenient evidence for a fact finder to determine compliance with the law.").

[275] *See* JESSICA FJELD, NELE ACHTEN, HANNAH HILLIGOSS, ADAM CHRISTOPHER NAGY &

Constitutional AI can help promote public understanding and engagement with the complex issues surrounding AI authority. By making the constitution-drafting process itself a focus of public discourse and debate, Public Constitutional AI can also foster a shared sense of ownership and investment in the principles that will shape the future of AI.[276] Citizens can come to see the AI constitution not as an esoteric technical document but as a living expression of their collective values and commitments.

The case law grounding approach can enhance the transparency and legitimacy of Public Constitutional AI even further.[277] By explicitly anchoring the interpretation and application of constitutional principles in specific cases and precedents, Public Constitutional AI can provide a more concrete and publicly accessible framework for evaluating the behavior of AI systems. The development of a rich body of AI common law through public deliberation and adjudication can help clarify the meaning and implications of abstract constitutional values, making them more relatable and actionable for both developers and citizens alike.[278] In

---

MADHULIKA SRIKUMAR, BERKMAN KLEIN CTR. FOR INTERNET & SOC'Y, PRINCIPLED ARTIFICIAL INTELLIGENCE: MAPPING CONSENSUS IN ETHICAL AND RIGHTS-BASED APPROACHES TO PRINCIPLES FOR AI 11–12 (2020) (discussing the purpose of AI and pointing out the role of principle articulation in defining comprehensive governance norms); *see also* Dennis Redeker, Lex Gill & Urs Gasser, *Towards Digital Constitutionalism? Mapping Attempts to Craft an Internet Bill of Rights*, 80 INT'L COMMC'N GAZETTE 302, 316 (2018) ("[I]nstead of dealing with the nuts and bolts of drafting legislation, digital constitutionalism as a societal constitutionalism allows to think about the 'ideal' Internet. The discourse thus provides an opportunity for action-oriented political philosophizing, having to collectively weight often-conflicting rights and principles. Even if non-binding, these documents can be influential, as they become a rallying point for civil society, media and politicians, and judges can take them as a point of orientation where no legislation exists." (citations omitted)).

[276] *See* Waldman, *supra* note 274, at 630 ("We need a robust, substantive approach to ensure that algorithmic systems meet fundamental social values other than efficiency. To do that, we need to audit the code of automated systems for noncompliance with values like equality, nondiscrimination, dignity, privacy, and human rights. Academic researchers have been doing this for some time and can be deputized to conduct independent sociotechnical analyses of algorithmic systems before and after they are used in commerce or by a government entity.").

[277] *See* Chen & Zhang, *supra* note 32, at 21 ("Making precedent selection an independent component also means we can understand the quality of any automated systems through evaluations and metrics, as well as gain insight into when existing precedents are insufficient.").

[278] *See* Margot E. Kaminski & Gianclaudio Malgieri, *Algorithmic Impact Assessments*

the same sense, the more the AI Court becomes salient and legitimate as an institution, the more legitimacy it lends to AI decisionmaking.[279]

This public engagement and understanding, in turn, can help mitigate the opacity of AI systems at a deeper level than mere publication of the constitution itself. If citizens have been actively involved in shaping the principles and values underlying AI governance, they may be better equipped to evaluate the legitimacy of specific AI systems and to hold developers and deployers, both private and public, accountable for adhering to those principles.[280] The opacity of individual decisions may be less daunting if there is a shared public understanding of the broader framework in which those decisions are made.

While increased public engagement and understanding fostered by Public Constitutional AI can help alleviate the systemic opacity of AI decisionmaking, its true transformative potential lies in its ability to imbue AI systems with a sense of democratic legitimacy rooted in popular authorship and contextualized human judgment.

*2. Political Community.* In modern democratic societies, the legitimacy of law, and especially of constitutional law, is deeply rooted in the idea of popular authorship.[281] As many scholars have

---

*Under the GDPR: Producing Multi-Layered Explanations*, 11 INT'L DATA PRIV. L. 125, 133 (2021) ("Public-facing disclosure enables public feedback, both in the form of market feedback (enabling individuals to avoid companies with bad policies) and in the form of regulatory feedback over the longer term (enabling individuals to elect representatives who will put in place laws that will prevent bad company behaviour).").

[279] For an article proving the "legality of intelligent judicial operation simultaneously from the four-dimensional perspectives of [AI's] intervention in judicial decision-making," see generally Zichun Xu, *The Legitimacy of Artificial Intelligence in Judicial Decision Making: Chinese Experience*, 13 INT'L J. TECHNOETHICS, no. 2, 2022, at 1.

[280] *See* Katyal, *supra* note 34, at 112 ("[Accountability algorithms] do more than emphasize transparency of authority. They provide important variables to consider in ensuring accuracy and auditability—urging designers to carefully investigate areas of error and uncertainty by undertaking sensitivity analysis, validity checks, and a process of error correction, and also enabling public auditing, if possible, or auditing by a third party, if not possible.").

[281] *See* Hanna Fenichel Pitkin, *The Idea of a Constitution*, 37 J. LEGAL EDUC. 167, 169 (1987) ("[A]lthough constituting is always a free action, how we are able to constitute ourselves is profoundly tied to how we are already constituted by our own distinctive history. Thus there is a sense, after all, in which our constitution is sacred and demands our respectful acknowledgement. If we mistake who we are, our efforts at constitutive action will fail."); ACKERMAN, *supra* note 112, at 6 ("Before gaining the authority to make supreme law in the name of the People, a movement's political partisans must, first, convince an extraordinary

observed, the law derives its authority from the fact that it is perceived to be a creation of the people themselves.[282] The U.S. Constitution, in particular, is seen as "an expression of the deepest beliefs and convictions of the American nation,"[283] and it sustains its legitimacy through the "quintessentially democratic attitude in which citizens know themselves as authorities, as authors of their own law."[284] When this works, the culture of law "obliges both individuals and groups through their words and deeds to take ownership of and make connections with a particular legal regime as facets of themselves."[285] It is through this active participation in the creation and interpretation of constitutional meaning that citizens come to see the law as an expression of their collective will and values.

Public Constitutional AI has several key advantages when it comes to mitigating the political community legitimacy deficit. First, Public Constitutional AI can be a product of a specific political community, reflecting not just universal values but the particular values and commitments of the people it serves. Unlike private AI systems, which are often developed with a global market in mind,[286] Public Constitutional AI would be grounded in the distinctive cultural, historical, and political context of the community that creates it. This rootedness in a particular democratic polity could help ensure that the AI system is seen as legitimate and responsive to the needs and interests of its constituents. This means that we can expect very different constitutional documents in different

---

number of their fellow citizens to take their proposed initiative with a seriousness that they do not normally accord to politics . . . ."); Post & Siegel, *supra* note 211, at 1982–83 ("In the American tradition, the authority of the Constitution is sustained through attitudes of veneration and deference, but it is also sustained through the quintessentially democratic attitude in which citizens know themselves as authorities, as authors of their own law."); Judith Resnik, *Law as Affiliation: "Foreign Law," Democratic Constitutionalism, and the Sovereigntism of the Nation-State*, 6 INT' J. CONST. L. 33, 35 (2008) ("Sovereigntism has taught us that proclaiming that 'our' law is uniquely ours has popular appeal; to do so speaks to the human aspiration to mark oneself as belonging to a group whose other members share affiliations, ideas, precepts, practices, and values.").

[282] *See* materials cited *supra* note 281.

[283] Post, *supra* note 112, at 36.

[284] Post & Siegel, *supra* note 211, at 1983.

[285] Resnik, *supra* note 281, at 35.

[286] *See, e.g.*, *About*, OPENAI, https://openai.com/about/ (last visited Apr. 2, 2025) ("Our mission is to ensure that artificial intelligence benefits all of humanity.").

jurisdictions—potentially tracking the diversity of constitutional law generally.[287]

Second, Public Constitutional AI would be a part of the public sphere and public discourse, rather than a mere product of a market-based entity. It would not ask citizens to simply trust big tech companies to do what is best for them but would instead allow the public to actively shape the development and deployment of AI through democratic processes. This could help foster greater public understanding and trust in AI systems, as well as provide a mechanism for holding them accountable to the values and priorities of the community.

Third, and perhaps most fundamentally, Public Constitutional AI can begin to address the judgment issue that undermines the legitimacy of AI decisionmaking. As discussed earlier, the legitimacy of law in a democratic society relies on the notion that legal judgments are the product of a dialectical relationship between the community and its representatives.[288] Judges and juries are seen as embodying the shared values and common sense of the community, and their decisions are validated through a process of public discourse and reflection.[289]

AI, however, lacks the capacity to engage in this kind of reflective judgment grounded in a shared communal context; its decisionmaking is based on the calculational processing of data and algorithms, which, however sophisticated, cannot replicate the temporally and culturally situated nature of human judgment.[290] By making the development and training of AI systems a part of the public discourse and deliberation, Public Constitutional AI could help bridge this gap. The principles and values that guide AI decisionmaking would therefore not be derived solely from the aggregation of data points but also from the shared understandings and commitments of the political community.

---

[287] *See* Nicola Palladino, *The Role of Epistemic Communities in the "Constitutionalization" of Internet Governance: The Example of the European Commission High-Level Expert Group on Artificial Intelligence*, 45 TELECOMMS. POL'Y, no. 6, 2021, at 1, 13 (highlighting the impact of diverse stakeholders on ethical guidelines).

[288] *See supra* Part II.A.

[289] *See, e.g.*, Post *supra* note 19, at 48 ("Judgments are validated by the reciprocal relationship between a community and its members . . . ." (footnote omitted)).

[290] *See supra* Part II.B; *see also* Stader, *supra* note 126, at 24–25 ("Reflecting judgement, for example in the context of ethical theories, must take into account the ambiguities of the theory, its conditional nature, and the purpose of judgement itself. This awareness of its limitations, tentativeness and incompleteness is the strength of judgement, because it leaves room for the resonance of reality and experience through time.").

Moreover, Public Constitutional AI represents a way of keeping humans "in the loop," even if not directly involved in every individual AI decision.[291] The public's participation in the drafting of the AI constitution and the ongoing development of AI case law through democratic processes ensures a form of indirect human oversight and input, and the AI system's decisions are thus grounded in human judgment at a foundational level—even if humans are not reviewing each specific output.[292] This is reinforced by the iterative process of training new models on the updated basis of an AI Court case law repository, allowing for the continued infusion of human values and understanding into the system.[293] In this way, Public Constitutional AI seeks to transform AI decisionmaking from something external and alien, to which we are subjected, into a system created and evolved through the direct intervention of the people (in the constitution-making stage) and their representatives (in the case law building stage). It represents a step towards making AI more transparent, accountable, and responsive to the shared judgments of democratic citizens, even if it

---

[291] *See generally* Kiel Brennan-Marquez & Stephen E. Henderson, *Artificial Intelligence and Role-Reversible Judgment*, 109 J. CRIM. & CRIMINOLOGY 137, 146–48 (2019) (noting the widespread view that "we ought to keep humans 'in the loop'" regarding the decisionmaking process).

[292] *See* Deven R. Desai & Joshua A. Kroll, *Trust but Verify: A Guide to Algorithms and the Law*, 31 HARV. J.L. & TECH. 1, 16–17 (2017) ("Although the private sector is regulated differently than the public sector, calls for transparency as it relates to software-based decision-making in the private sector abound. . . . In the same vein, other studies and investigations have identified a range of examples where software was part of undesired or troubling outcomes and have called for methods to detect such issues.").

[293] *See* Shlomit Yanisky-Ravid & Sean K. Hallisey, *"Equality and Privacy by Design": A New Model of Artificial Intelligence Data Transparency Via Auditing, Certification, and Safe Harbor Regimes*, 46 FORDHAM URB. L.J. 428, 472–73 (2019) ("[U]sing specific types of data, or using data in specific contexts, can create issues that implicate privacy concerns that trainers of AI systems ought to consider. Whether it is the type of data, such as health data, or the subject of data, such as children, or the context in which data is used, such as in consumer leading or other economic sectors protected by anti-discrimination legislation, AI systems are operating in spaces governed by existing privacy and anti-discrimination regulatory regimes. . . . AI stakeholders, including operators, must ensure that the datasets used to train their AI systems do not violate these regulatory regimes.").

cannot entirely replicate the richness and nuance of human decisionmaking in each individual case.[294]

This is not to suggest that Public Constitutional AI can fully resolve the judgment issue or imbue AI with the same capacity for contextual understanding and norm-creation as humans. After all, the opacity and scale of AI systems may always pose challenges for democratic legitimacy.[295] However, by subjecting the development of AI to public scrutiny and debate, and by grounding its decisionmaking in publicly articulated values and principles, Public Constitutional AI could help to create a stronger connection between AI and the communities it serves.

Finally, Public Constitutional AI represents a new form of governance that combines elements of both private and public control, offering a promising path for achieving a more balanced distribution of power between the state, the market, and civil society in the development and deployment of AI systems. Rather than a fully state-controlled approach, which risks concentrating excessive authority in government hands and stifling private innovation,[296] Public Constitutional AI envisions a collaborative governance model in which private companies, research institutions, and other nonstate actors work together with democratic institutions to create AI systems that serve the common good. This kind of hybrid governance has proven successful in other domains, such as the development of the Internet,[297] where a "multistakeholder" approach has helped to preserve a degree of decentralization and openness while still allowing for public oversight and coordination.[298] By leveraging the expertise and creativity of the private sector within a framework of democratic accountability, Public Constitutional AI could help to ensure that AI development can be seen as responsive to the needs and values of the broader political community while still harnessing the immense innovative potential of private enterprise.[299]

---

[294] *See generally* Brennan-Marquez & Henderson, *supra* note 291, at 152–56 (discussing the limitations of AI in replicating human judgment and the need for human oversight).

[295] *See* Stader, *supra* note 126, at 25 ("The structure of judgement, linking something general to something particular, allows it to have reasons, to justify itself. [AI] cannot give reasons, because it does not have reasons in the way a judgement has, it has data and statistical calculation. It does not refer to a constantly changing lifeworld, but to a present data set, which it calculates iteratively.").

[296] *See, e.g.*, Jack M. Balkin, *Free Speech in the Algorithmic Society: Big Data, Private Governance, and New School Speech Regulation*, 51 U.C. DAVIS L. REV. 1149, 1194–96 (2018) (noting the risks of state overreach in the regulation of digital technologies).

[297] *See, e.g.*, Lawrence B. Solum, *Models of Internet Governance*, *in* INTERNET GOVERNANCE: INFRASTRUCTURE AND INSTITUTIONS 48, 57–58 (Lee A. Bygrave & Jon Bing eds., 2009) (describing the necessity of a hybrid model of Internet governance).

In sum, the Public Constitutional AI approach offers hope for addressing two critical issues in the AI legitimacy crisis: the opacity deficit and the political community deficit. Involving citizens in creating and interpreting an AI constitution grounds AI development in the participatory processes and shared values of a democratic society. This imbues AI systems with popular authorship and human judgment embedded in a particular social and political context.

## V. CONCLUSION

The transformative potential of AI authority, in both the private and public spheres, is already upon us. Yet our institutions are only beginning to grapple with the profound challenges to democratic legitimacy that these technologies present. While still in the early days of the development and implementation of AI, the breathtaking speed of innovation means that we cannot afford to wait for the technology to take its mature shape before integrating it into our political and legal frameworks.

Public Constitutional AI is meant to start this essential conversation. By imagining a future where AI decisionmaking is grounded in participatory processes, public deliberation, and the collective will of the communities it serves, we can begin to chart a course towards a legal and political context in which AI is not an alien force but a legitimate expression of the people. Though the precise form of this approach will undoubtedly evolve through trial and error, the fundamental insight—AI can be tethered to the public through constitutional principles—provides a guiding light for the road ahead.

---

[298] *See, e.g.,* Milton Mueller, John Mathiason & Hans Klein, *The Internet and Global Governance: Principles and Norms for a New Regime*, 13 GLOB. GOVERNANCE 237, 246 (2007) ("[T]he Internet allows for the privatization and decentralization of network operations and policies; it also facilitates privatization and decentralization of software applications and the ability to originate information content as well. This principle means that the Internet has less need than many other systems for global governance.").

[299] *See supra* note 14 and accompanying text.